\documentclass[aps,prb,superscriptaddress,longbibliography,reprint]{revtex4-1}

\usepackage[utf8]{inputenc}
\usepackage{graphicx,dcolumn,bm,amsmath,amssymb,euscript}

	\graphicspath{{figures/}}

\newcommand{\kapt}{\tilde{\kappa}}
\newcommand{\kapsq}{\langle \kappa^2 \rangle}
\newcommand{\rhot}{\tilde{\rho}}
\newcommand{\pt}{\tilde{p}}
\newcommand{\qt}{\tilde{q}}

	\usepackage{color, colortbl}   
	\usepackage[bookmarksopen,bookmarksnumbered,colorlinks,citecolor=red,urlcolor=blue,filecolor=green]{hyperref}

\begin{document}

\title{Dynamics of Atomic Steps on GaN (0001) during Vapor Phase Epitaxy}

\author{Guangxu Ju}
    \email[correspondence to: ]{juguangxu@gmail.com}
	\altaffiliation[current address: ]{Lumileds Lighting Co., San Jose, CA 95131 USA.}
	\affiliation{Materials Science Division, Argonne National Laboratory, Argonne, IL 60439 USA}
\author{Dongwei Xu}
	\affiliation{Materials Science Division, Argonne National Laboratory, Argonne, IL 60439 USA}
	\affiliation{School of Energy and Power Engineering, Huazhong University of Science and Technology, Wuhan, Hubei 430074, China}
\author{Carol Thompson}
	\affiliation{Department of Physics, Northern Illinois University, DeKalb, IL 60115 USA}
\author{Matthew J. Highland}
	\affiliation{X-ray Science Division, Argonne National Laboratory, Argonne, IL 60439 USA}
\author{Jeffrey A. Eastman}
	\affiliation{Materials Science Division, Argonne National Laboratory, Argonne, IL 60439 USA}
\author{Weronika Walkosz}
	\affiliation{Department of Physics, Lake Forest College, Lake Forest, IL 60045 USA}
\author{Peter Zapol}
	\affiliation{Materials Science Division, Argonne National Laboratory, Argonne, IL 60439 USA}
\author{G. Brian Stephenson}
    \email[correspondence to: ]{stephenson@anl.gov}
	\affiliation{Materials Science Division, Argonne National Laboratory, Argonne, IL 60439 USA}

\date{July 5, 2020}  

\begin{abstract}
Images of the morphology of GaN (0001) surfaces often show half-unit-cell-height steps separating a sequence of terraces having alternating large and small widths. This can be explained by the $\alpha \beta \alpha \beta$ stacking sequence of the wurtzite crystal structure, which results in steps with alternating $A$ and $B$ edge structures for the lowest energy step azimuths, i.e. steps normal to $[0 1 \overline{1} 0]$ type directions. Predicted differences in the adatom attachment kinetics at $A$ and $B$ steps would lead to alternating $\alpha$ and $\beta$ terrace widths. However, because of the difficulty of experimentally identifying which step is $A$ or $B$, it has not been possible to determine the absolute difference in their behavior, e.g. which step has higher adatom attachment rate constants. Here we show that surface X-ray scattering can measure the fraction of $\alpha$ and $\beta$ terraces, and thus unambiguously differentiate the growth dynamics of $A$ and $B$ steps. We first present calculations of the intensity profiles of GaN crystal truncation rods (CTRs) that demonstrate a marked dependence on the $\alpha$ terrace fraction $f_\alpha$. We then present surface X-ray scattering measurements performed \textit{in situ} during homoepitaxial growth on (0001) GaN by vapor phase epitaxy. By analyzing the shapes of the $(1 0 \overline{1} L)$ and $(0 1 \overline{1} L)$ CTRs, we determine that the steady-state $f_\alpha$ increases at higher growth rate, indicating that attachment rate constants are higher at $A$ steps than at $B$ steps. We also observe the dynamics of $f_\alpha$ after growth conditions are changed. The results are analyzed using a Burton-Cabrera-Frank model for a surface with alternating step types, to extract values for the kinetic parameters of $A$ and $B$ steps. These are compared with predictions for GaN (0001). 
\end{abstract}

\maketitle
\section{Introduction}

The atomic-scale mechanisms of crystal growth are often described within the framework of Burton-Cabrera-Frank (BCF) theory \cite{1951_Burton_PhilTransRS243_29,1999_Jeong_SurfSciRep34_171,2015_Woodruff_PhilTransA373_20140230}, in which atoms are added to the growing crystal surface by preferential attachment at the steps forming the edges of each exposed atomic layer, or terrace.
The motion of the steps during growth defines the classical homoepitaxial crystal growth modes of 1-dimensional step flow, 2-dimensional island nucleation and coalescence, or 3-dimensional roughening \cite{1993_Tsao_MatFundMBE}.
The BCF model was originally developed for crystals with simple symmetries, with step heights of a full unit cell and step properties that are identical from step to step, for a fixed step direction (in-plane azimuth).
However, when the space group of the crystal includes screw axes or glide planes,
the growth behavior of steps on surfaces perpendicular to one of these symmetry elements can have fundamentally different characteristics \cite{2004_vanEnckevort_ActaCrystA60_532}.
In this case, each succeeding terrace has the same atomic termination, but a different in-plane orientation.
These terraces are separated by fractional-unit-cell-height steps.
Even for a fixed step azimuth the step structures and properties can vary from step to step.
A well-studied example is the Si (001) surface, which is normal to a $4_2$ screw axis in the diamond cubic structure.
Because the surface reconstructs strongly and reveals the orientation of each terrace, the two types of steps can be clearly identified \cite{1985_Tromp_PRL55_1303}.

\begin{figure}
\includegraphics[width=1\linewidth]{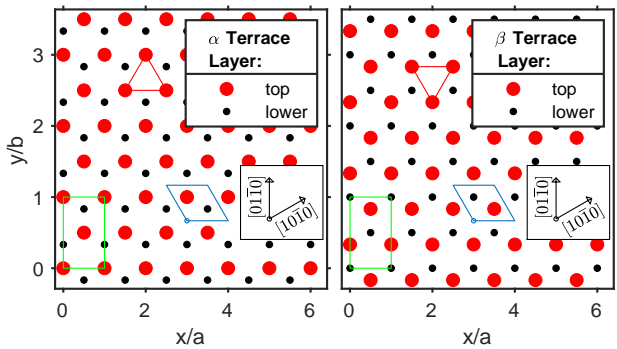}
\caption{Structure of $\alpha$ and $\beta$ terraces of the (0001) surface of an HCP-type crystal, e.g. the Ga sites in wurtzite-structure GaN. Red triangle of top-layer sites around $6_3$ screw axis shows difference between alternating $\alpha$ and $\beta$ layers. Blue rhombus shows conventional HCP unit cell; green rectangle shows orthohexagonal unit cell. Axes give coordinates in terms of orthohexagonal lattice parameters $a$ and $b = \sqrt{3} a$. \label{fig:alpha_beta}}
\end{figure}

\begin{figure}
\includegraphics[width=0.8\linewidth]{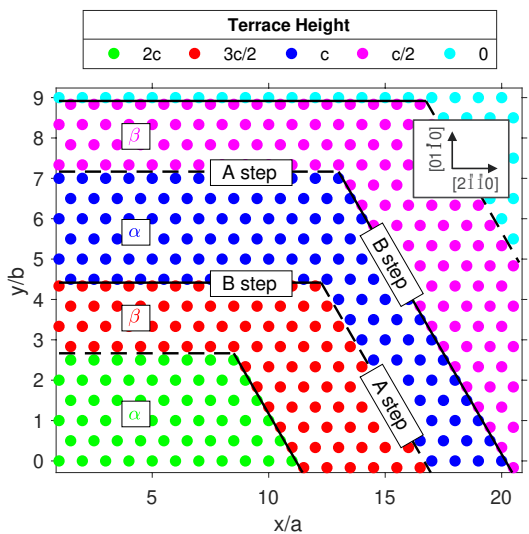}
\caption{Terrace and step structure of vicinal (0001) surface of an HCP-type crystal. Circles show in-plane positions of top-layer atoms on each terrace, with color indicating height. Orthohexagonal lattice parameters are $a$, $b$, and $c$. Steps of height $c/2$ typically have lowest edge energy when they are normal to $[0 1 \overline{1} 0]$, $[1 0 \overline{1} 0]$, or $[1 \overline{1} 0 0]$. Steps of a given azimuth have alternating structures, $A$ and $B$. The step structure changes from $A$ to $B$ or $B$ to $A$ when they change azimuth by $60^{\circ}$.
\label{fig:A_B}}
\end{figure}

A more subtle and widespread version of this effect occurs on the basal-plane $\{0001\}$-type surfaces of crystals having hexagonal close-packed (HCP) or related structures, which are normal to a $6_3$ screw axis.
The close-packed layers in HCP crystals have 3-fold symmetry alternating between $180^\circ$-rotated orientations from layer to layer, as shown by the $\alpha$ and $\beta$ terrace structures in Fig.~\ref{fig:alpha_beta}.
The $\alpha \beta \alpha \beta$ stacking sequence typically results in half-unit-cell-height steps on vicinal surfaces.
Often the lowest energy steps are normal to $[0 1 \overline{1} 0]$-type directions.
The alternating structures of these steps are conventionally labelled $A$ and $B$ \cite{1999_Xie_PRL82_2749,2001_Giesen_ProgSurfSci68_1}
as shown on Fig.~\ref{fig:A_B}.
When the in-plane azimuth of an $A$ step changes by $60^\circ$, e.g. from $[0 1 \overline{1} 0]$ to $[1 0 \overline{1} 0]$, its structure changes to $B$, and \textit{vice versa}.
Differences in the dynamics of adatom attachment at $A$ and $B$ steps have strong effects on the surface morphology produced during growth.

\begin{figure}
\includegraphics[width=0.5\linewidth]{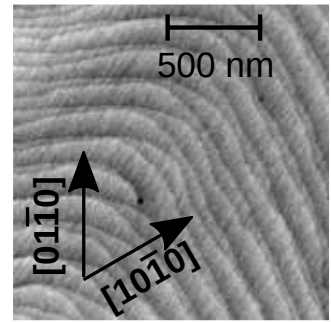} 
\caption{AFM height image of GaN (0001) surface typical of films grown on sapphire substrates by OMVPE, showing
regions of alternating step spacings and interlacing at corners where step azimuth changes by $\sim 60^\circ$. Step heights are c/2 (2.6 \AA). \label{fig:AFM_film}}
\end{figure}

Images of $\{0001\}$ surfaces showing the alternating nature of the steps have been obtained for several HCP-related systems, including SiC \cite{1951_Verma_PhilMag42_1005,1979_Sunagawa_JCrystGrowth46_451,1982_vanderHoek_JCrystGrowth58_545}, GaN \cite{1999_Xie_PRL82_2749,1999_Heying_JAP85_6470,2001_Vezian_MatSciEngB82_56,2002_Zauner_JCrystGrowth240_14,2006_Xie_PRB_085314,2007_Krukowski_CrystResTechnol42_1281,2008_Zheng_PRB77_045303,2013_Turski_JCrystGrowth367_115,2013_Lin_ApplPhysExp6_035503}, AlN \cite{2017_Pristovsek_physstatsolb254_1600711}, and ZnO \cite{2002_Chen_APL80_1358}. 
As shown in Fig.~\ref{fig:AFM_film}, such images typically indicate a tendency for local pairing of steps (i.e. alternating step spacings), and an ``interlaced'' structure in which the step pairs switch partners at corners where the step azimuth changes by $60^\circ$.
In some cases of MBE-grown GaN \cite{1999_Xie_PRL82_2749,2001_Vezian_MatSciEngB82_56,2013_Turski_JCrystGrowth367_115}, every other step takes on a zigzag morphology, so that all steps are made of segments of only one type, $A$ or $B$.
Similar alternating straight and crenellated steps have been observed in OMVPE of AlN \cite{2017_Pristovsek_physstatsolb254_1600711}.
Observations of triangular islands \cite{1999_Xie_PRL82_2749,2006_Xie_PRB_085314,2008_Zheng_PRB77_045303} indicate that one step may grow faster, leaving behind island shapes terminated by the slower growing step.
All of these features are consistent with predictions that $A$ and $B$ steps can have significantly different energies and/or attachment kinetics \cite{1999_Xie_PRL82_2749,2006_Xie_PRB_085314,2013_Turski_JCrystGrowth367_115,2011_Zaluska-Kotur_JAP109_023515,2010_Zaluska-Kotur_JNoncrystSolids356_1935,2017_Xu_JChemPhys146_144702,2017_Chugh_ApplSurfSci422_1120,2020_Akiyama_JCrystGrowth532_125410,2020_Akiyama_JJAP59_SGGK03}.
In particular, different attachment kinetics at $A$ and $B$ steps can produce a tendency to step pairing during growth and thus to different local fractions of $\alpha$ and $\beta$ terraces.
In limiting cases, the $\alpha$ terrace fraction $f_\alpha$ can approach zero or unity, when pairs of half-unit-cell-height steps join to form full-unit-cell-height steps.
However, in contrast to Si (001), for $\{0001\}$ surfaces of HCP-related systems it has been difficult to distinguish experimentally the terrace orientation, and thus to determine whether a given set of steps is of $A$ or $B$ type.

The dynamic properties of $A$ and $B$ steps on GaN (0001) have been predicted in several publications.
A seminal study \cite{1999_Xie_PRL82_2749} of MBE growth of GaN posited a higher tendency for adatom attachment at $A$ steps than at $B$ steps, giving faster $A$ steps for a given supersaturation.
The support for this result is based on an argument regarding the difference in dangling bonds between $A$ and $B$ steps, and a comparison with experimental results on GaAs (111) \cite{1997_Avery_PRL79_3938,2012_Jo_CrystGrowthDes12_1411}. 
Face-centered cubic (FCC) materials such as GaAs have $A$ and $B$ type steps on (111) surfaces that do not alternate between successive terraces and thus can be distinguished by their orientation \cite{2001_Giesen_ProgSurfSci68_1}.
In contrast, subsequent theoretical studies of GaN (0001) have consistently predicted that $A$ steps have \textit{smaller} adatom attachment coefficients than $B$ steps.
Kinetic Monte Carlo (KMC) studies of GaN (0001) growth under organo-metallic vapor phase epitaxy (OMVPE) conditions found step pairing \cite{2011_Zaluska-Kotur_JAP109_023515} driven by faster kinetics at $B$ steps than $A$ steps \cite{2010_Zaluska-Kotur_JNoncrystSolids356_1935}.
A KMC study of growth on an HCP lattice \cite{2017_Xu_JChemPhys146_144702} found a much lower Ehrlich-Schwoebel (ES) barrier at $B$ steps than at $A$ steps, when only nearest-neighbor jumps are allowed.
A recent KMC study of GaN (0001) growth under MBE conditions \cite{2017_Chugh_ApplSurfSci422_1120} found triangular islands that close analysis reveals are bounded by $A$ steps, indicating faster growth of $B$ steps.
An analysis of InGaN (0001) growth by MBE \cite{2013_Turski_JCrystGrowth367_115} concluded that adatom attachment at $B$ steps is faster, converting them into crenelated edges terminated by $A$ steps.
\textit{Ab initio} calculations of kinetic barriers at steps under MBE conditions \cite{2020_Akiyama_JCrystGrowth532_125410,2020_Akiyama_JJAP59_SGGK03} found a negative ES barrier at $B$ steps and a high positive ES barrier at $A$ steps, and a Ga attachment energy of $-0.78$ or $+1.27$~eV for $B$ or $A$ steps, indicating that attachment of Ga adatoms is preferable at $B$ steps.

The differences in these predictions reflect different assumptions about the growth environment, which we expect will affect the dynamics of $A$ and $B$ steps.
An experimental study of AlN (0001) surfaces grown by OMVPE \cite{2017_Pristovsek_physstatsolb254_1600711} found a change in the terrace fraction as a function of the V/III ratio used during growth.
Studies of islands on the FCC Pt (111) surface \cite{1998_Kalff_PRL81_1255,2009_Yin_APL94_183107} have found that $A$ steps have a higher growth rate than $B$ steps, but that this relationship is reversed by the presence of adsorbates such as CO.
Thus there is a clear need for a method for experimental determination of the difference between adatom attachment kinetics at $A$ and $B$ steps, especially an \textit{in situ} measurement in the relevant growth environment.

Here we demonstrate the use of \textit{in situ} surface X-ray scattering to distinguish the fraction of the surface covered by $\alpha$ or $\beta$ terraces during growth, and thus unambiguously determine differences in the attachment kinetics at $A$ and $B$ steps.
This development is made possible by the use of micron-scale X-ray beams and high-quality single-crystal substrates to investigate surface regions that have fixed step azimuths.
We first develop expressions for the surface scattering intensities that allow determination of terrace fraction $f_\alpha$, including the effect of surface reconstruction.
We then present measurements of $(0 1 \overline{1} L)$ and $(1 0 \overline{1} L)$ crystal truncation rods carried out \textit{in situ} during growth of GaN on the (0001) surface via OMVPE.
We fit these to obtain the variation of the steady-state $f_\alpha$ as a function of growth conditions, as well as the relaxation times of $f_\alpha$ upon changing conditions.
These results are compared to calculated dynamics based on an extension of the BCF model to systems with alternating step types, to quantify the differences in the attachment rates at $A$ and $B$ steps for GaN growth by OMVPE.

\section{Surface X-ray Scattering Theory}

In this section we develop expressions for the intensity distributions along crystal truncation rods for the GaN (0001) surface and demonstrate how they are sensitive to the fraction of surface covered by $\alpha$ or $\beta$ terraces.
Crystal truncation rods (CTRs) are streaks of scattering intensity extending in reciprocal space away from every Bragg peak in the direction normal to the crystal surface, due to the truncation of the bulk crystal \cite{1986_Robinson_PRB33_3830}.
For a vicinal surface, the CTRs are tilted away from the crystal axes, so that the CTRs from different Bragg peaks do not overlap.
The intensity distribution along a CTR is sensitive to the surface structure.
Here we include the effect of surface reconstruction, using relaxed atomic coordinates that have been calculated previously \cite{2012_Walkosz_PRB_85_033308}.

For these calculations it is convenient to introduce an orthohexagonal coordinate system \cite{1965_Otte_PhysStatSol9_441} with an orthorhombic unit cell having orthogonal in-plane lattice parameters $a$ and $b \equiv a \sqrt{3}$, where $a$ is the in-plane lattice parameter of the conventional hexagonal unit cell, as shown in Fig.~\ref{fig:alpha_beta}.
The out-of-plane lattice parameter $c$ is the same in both coordinate systems.
This gives Cartesian $x$, $y$, and $z$ axes parallel to the $[2 \overline{1} \overline{1} 0]$, $[0 1 \overline{1} 0]$, and [0001] directions, respectively.
In reciprocal space, the orthohexagonal coordinates $H^\prime K^\prime L^\prime$ are related to the standard hexagonal Miller-Bravais reciprocal space coordinates $H K I L$ by $H^\prime = H$, $K^\prime = H+2K$, $L^\prime = L$.
Thus the Cartesian components of the scattering wavevector $\mathbf{Q}$ are given by $Q_x = (2 \pi / a) H$, $Q_y = (2 \pi / b) (H + 2K)$, and $Q_z = (2 \pi / c) L$.
When referring to Bragg peaks, planes, etc., we will continue to use standard hexagonal Miller-Bravais indices $HKIL$.

The X-ray reflectivity along the CTRs can be calculated by adding the complex amplitudes from the substrate crystal and the reconstructed overlayers, with proper phase relationships.
We start with the simple case of an exactly-oriented (0001) surface without steps, and then extend this to the case of a vicinal surface having an array of straight steps, as in previous work \cite{1999_Munkholm_JApplCryst32_143,2002_Trainor_JApplCryst35_696}.
We neglect effects of refraction when the incident or exit beams are near the critical angle, and for calculating absorption effects we assume the incident and exit angles with respect to the surface are equal.

\subsection{Exactly oriented surface with reconstruction}

For an exactly oriented (0001) surface, the CTRs extend continuously in the $Q_z$ direction at fixed $Q_x = (2 \pi/a) H_0$ and $Q_y = (2 \pi/b) (H_0 + 2K_0)$ through each Bragg peak $H_0 K_0 I_0 L_0$, where these indices are integers.
The CTRs thus connect all the Bragg peaks of different $L_0$ at the same $H_0 K_0 I_0$.

The contribution to the complex amplitude of the reflectivity from the truncated crystal substrate below the reconstructed overlayers is
\begin{equation}
    r_s = r_f F_s \sum_{\ell = -\infty}^0 Z^\ell = r_f F_s \frac{Z}{Z - 1},
    \label{eq:r_s}
\end{equation}
where $r_f \equiv 4 \pi i r_0 / (2 a b Q)$, $r_0 = 2.817 \times 10^{-13}$~cm is the Thomson radius of the electron, and $Q$ is the magnitude of the wavevector.  
The substrate structure factor $F_s$ is
\begin{equation}
    F_s = \sum_k f_k(Q) \exp(-\sigma_k^2 Q^2) \sum_n \exp(i \mathbf{Q} \cdot \mathbf{r}_{kn}^s).
\end{equation}
Here the first sum is over the chemical elements present in the crystal (in our case Ga and N), $f_k(Q)$ is the atomic form factor of element $k$, $\sigma_k$ is a Debye-Waller thermal vibration length for element $k$, the second sum is over the substrate atoms of type $k$ in a unit cell, and $\mathbf{r}_{kn}^s$ is the position of substrate atom $n$ of type $k$.
We consider Ga-face (0001) surfaces with a Ga termination for the substrate.
Since the atomic coordinates for the reconstructed overlayers were calculated using a $2 \times 2$ unit cell, for consistency the unit cell sums used in calculating the structure factors are carried out over two adjacent orthohexagonal unit cells having an area $2 a b$, which is normalized out in the denominator of $r_f$.
Table~\ref{tab:coord_sub} in Appendix B lists the atomic coordinates used.

The quantity $Z$ in Eq.~(\ref{eq:r_s}) is the ratio of the contribution from one unit cell to that from the unit cell at $\Delta z = -c$ below it.
It consists of a phase factor and an absorption factor,
\begin{equation}
    Z \equiv \exp (i Q_z c + \epsilon c / Q_z),
\end{equation}
where $\epsilon = 4 \pi / (\lambda \ell_{abs})$ is related to the photon wavelength $\lambda$ and absorption length $\ell_{abs}$.
One can see from Eq.~(\ref{eq:r_s}) that the scattering is built up by summing the contributions from each layer of the semi-infinite crystal in the $z$ direction from $\ell = -\infty$ to $\ell = 0$.

We consider reconstructions in which the Ga and N atoms in the top layer of unit cells at the surface are relaxed from their bulk crystal positions, and there can be extra Ga, N, and/or H atoms bonded to the surface \cite{2012_Walkosz_PRB_85_033308}.
The reflectivity from this reconstructed overlayer is
\begin{equation}
    r_r = r_f F_r Z, 
\end{equation}
where the structure factor of the reconstruction $F_r$ is
\begin{equation}
    F_r = \sum_j \theta_{j} \sum_k f_k \exp(-\sigma_k^2 Q^2) \sum_n \exp(i \mathbf{Q} \cdot \mathbf{r}_{jkn}).
\end{equation}
Here the first sum is over the 6 possible domain orientations of the reconstruction, $\theta_{j}$ is the fraction of domain $j$, the second sum is over the chemical elements present in the reconstruction (Ga, N, and H), the third sum is over the atoms of type $k$ in a unit cell, and $\mathbf{r}_{jkn}$ is the position of atom $n$ of type $k$ in domain orientation $j$.
The 6 domain orientations are related by 3-fold rotation about the $6_3$ axis, and/or reflection about a $(2 \overline{1} \overline{1} 0)$ plane passing through the axis (e.g. $x = 0$).
The total reflectivity amplitude is the sum of the complex amplitudes from the substrate and the reconstructed overlayer,
\begin{equation}
    r_t = r_s + r_r.
\end{equation}

The reflectivity amplitudes calculated above are for the kinematic  limit in which the reflectivity is much smaller than unity.
Near the Bragg peaks, where the reflectivity amplitude of the substrate approaches unity, the amplitude can be corrected using
\begin{equation}
    r_t^{dyn} = \frac{2 r_t}{1 + \sqrt{1 + 4 r_t^2}},
    \label{eq:dyn}
\end{equation}
which insures the reflectivity does not exceed unity.
The intensity reflectivity is the square of the modulus of the amplitude reflectivity,
\begin{equation}
    R = |r_t^{dyn}|^2 \exp[-\sigma_R^2 (Q_z - Q_z^B)^2],
    \label{eq:R}
\end{equation}
where the final factor has been introduced to account for surface roughness having an RMS value of $\sigma_R$, with $Q_z^B$ being the $Q_z$ of the nearest Bragg peak on the CTR.

\begin{figure}
\includegraphics[width=1\linewidth]{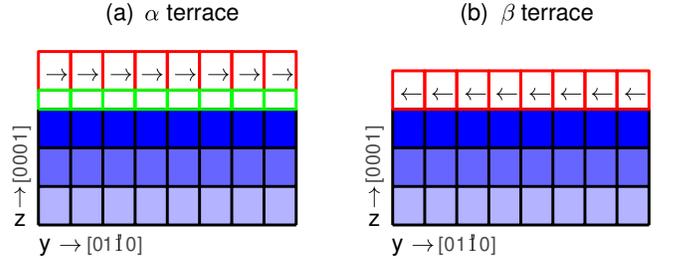}
\caption{Substrate unit cells (black) and reconstructed unit cells (red), for exactly oriented (0001) surface with (a) $\alpha$ and (b) $\beta$ termination.
In (a), the extra half unit cells producing the shift between $\alpha$ and $\beta$ are shown in green.
Blue shade indicates index $\ell$ of sum in Eq.~(\ref{eq:r_s}), with final term $\ell = 0$ darkest. \label{fig:exact}}
\end{figure}

To compare the scattering from surfaces terminated at $\alpha$ and $\beta$ terraces, we terminate the substrate at a $\beta$ terrace, and incorporate an extra half unit cell of substrate atoms (in their bulk positions) into the bottom of the reconstructed overlayer for the $\alpha$ terrace case.
We also reverse the relaxation amounts in the $y$ direction for the $\alpha$ terraces, relative to those for the $\beta$ terraces.
Figure~\ref{fig:exact} illustrates these arrangements.
Appendix B gives tables of atomic coordinates $\mathbf{r}_{jkn}$ used for the $\alpha$ and $\beta$ structure factors.

\begin{figure}
\includegraphics[width=1\linewidth]{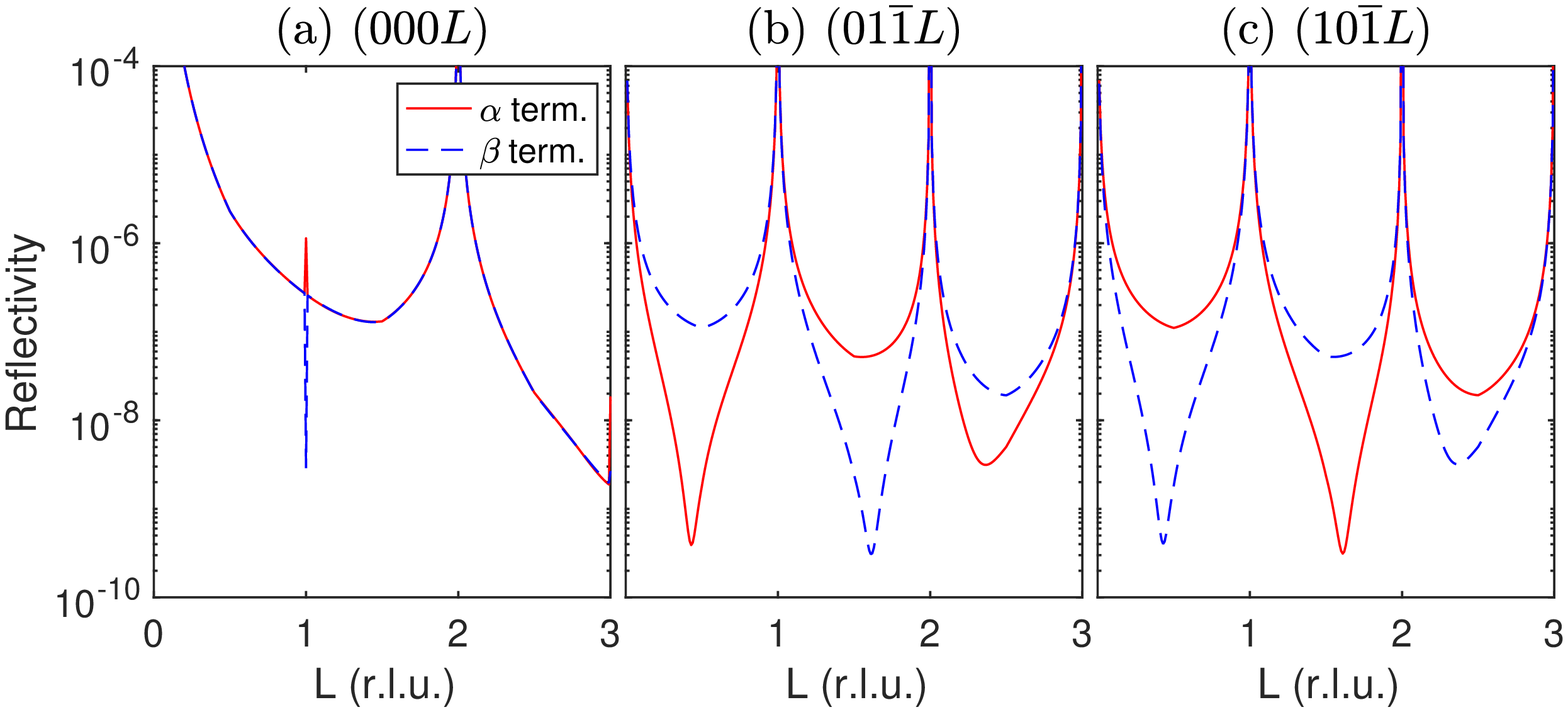}
\caption{Calculated reflectivity of (a) (000L), (b) $(0 1 \overline{1} L)$, and (c) $(1 0 \overline{1} L)$ CTRs for exactly oriented surface with $\alpha$ or $\beta$ terrace terminations, with the 3H(T1) reconstruction, $\sigma_R = 1$~\AA, $\sigma_k = 0.11$~\AA.
\label{fig:CTR_exact_3HT1}}
\end{figure}

Figure~\ref{fig:CTR_exact_3HT1} shows the calculated reflectivity as a function of $L$ for different integer $H_0 K_0 I_0$ values, for both $\alpha$ and $\beta$ terminations.
Fits to X-ray measurements described below indicate that the GaN surface under OMVPE conditions has a 3H(T1) reconstruction, in which 3 of every 4 Ga atoms in top-layer sites shown in Fig.~\ref{fig:A_B} is bonded to an adsorbed hydrogen.
We thus show calculations for a surface with the 3H(T1) reconstruction, for equal fractions $\theta_j = 1/6$ of all six domains.
We use atomic form factors for each type of atom \cite{1995_Waasmaier_ActaCrystA51_416} with resonant corrections for the 25.75 keV photon energy used in the experiments \cite{1993_Henke_ANDT54_181}, and an estimated Debye-Waller length of $\sigma_k = 0.11$~\AA~ for all atoms.
The (000L) CTR is insensitive to the difference between the $\alpha$ and $\beta$ terminations; both give the same intensity distribution.
In contrast, the $(0 1 \overline{1} L)$ and $(1 0 \overline1 L)$ CTRs show very different intensity distributions for $\alpha$ and $\beta$ terminations.
There are alternating deep and shallow minima between the Bragg peaks, with the alternation being opposite for the two terminations.
Furthermore, the $(0 1 \overline{1} L)$ scattering from the $\alpha$ terrace is identical to the $(1 0 \overline{1} L)$ scattering from the $\beta$ terrace, and vice versa, as required by symmetry.
We have performed calculations using atomic coordinates for all of the GaN (0001) reconstructions found previously \cite{2012_Walkosz_PRB_85_033308}, as well as an unreconstructed surface.
All show the same qualitative behavior, with small quantitative differences.
Furthermore, because the X-ray scattering is dominated by the Ga atoms, which occupy an HCP lattice, the same qualitative behavior is also obtained for an elemental HCP crystal.

\subsection{Vicinal surface with reconstruction}

\begin{figure}
\includegraphics[width=1\linewidth]{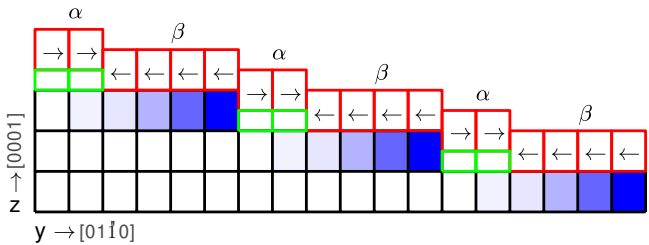}
\caption{Substrate unit cells (black), extra half unit cells (green) producing shift between $\alpha$ and $\beta$ terminations of neighboring terraces, and reconstructed unit cells (red), for a vicinal surface with $m = 6$, $m_\alpha = 2$, and $f_\alpha = 1/3$. Blue shade indicates index $\ell$ of sum in Eq.~(\ref{eq:r_s_v}), with final term $\ell = 0$ darkest. \label{fig:recon}}
\end{figure}

We now consider a vicinal surface, with a periodic array of steps.
We specialize to steps normal to the $[0 1 \overline{1} 0]$ $y$ axis.
We assume that the surface height decreases by a full unit cell $c$ every $m$ unit cells in $y$, so that the period of the step array is $mb$.
The surface offcut angle $\gamma$ relative to (0001) is given by $\tan{\gamma} = c / (mb)$, and the surface is parallel to $(0 1 \overline{1} 2m)$ planes.
The CTRs from this surface are tilted in the $Q_y$ direction at an angle $\gamma$ from (0001).
Because of the tilt, there are $2m$ times as many CTRs as in the exactly oriented case, indexed not just by $H_0 K_0 I_0$ but also by values of $L_0$ from $0$ to $2m-1$.
The $Q_y$ value varies with $L$ along the CTR according to $Q_y = (2 \pi / b) [H_0 + 2K_0 + (L-L_0)/m]$, where $H_0 K_0 I_0 L_0$ is the primary Bragg peak associated with the CTR.
The spacing in $L$ along a given CTR between Bragg peak positions is $2m$, rather than unity as in the exactly oriented surface.
Figure \ref{fig:recon} shows the substrate and reconstructed unit cells used to calculate the CTRs for the vicinal surface.
The width of the $\alpha$ terraces is $m_\alpha$ unit cells, and the width of the $\beta$ terraces is $m - m_\alpha$ unit cells.
The $\alpha$ terrace fraction is given by $f_\alpha = m_\alpha / m$.

The reflectivity amplitude from the truncated crystal substrate is
\begin{equation}
    r_s = \frac{r_f F_s}{m} \sum_{\ell = -\infty}^0 Y^\ell = \frac{r_f F_s}{m} \frac{Y}{Y - 1}, \label{eq:r_s_v} 
\end{equation}
where the quantity $Y$ is now the ratio of the contribution from one unit cell to that from the unit cell at $\Delta y = -b$ beside it,
\begin{equation}
    Y \equiv \exp (i Q_y b + \epsilon \, b \sin \gamma / Q_\perp),
\end{equation}
where $Q_\perp = Q_z/\cos{\gamma} + Q_y \sin{\gamma}$ is the component of $\mathbf{Q}$ perpendicular to the surface.
For a vicinal crystal, the scattering is built up by summing the contributions from each unit cell in the $y$ direction from $\ell = -\infty$ to $\ell = 0$. 

The reflectivity from the reconstructed layers on the $\alpha$ terraces can be written as
\begin{equation}
    r_\alpha = \frac{r_f F_\alpha}{m} \sum_{\ell = 1}^{m_\alpha} Y^\ell = \frac{r_f F_\alpha}{m} \frac{Y(Y^{m_\alpha} - 1)}{Y - 1},
    \label{eq:ra}
\end{equation}
where the unit cell structure factor $F_\alpha$ is given by
\begin{equation}
    F_\alpha = \sum_j \theta_{\alpha j} \sum_k f_k \exp(-\sigma_k^2 Q^2) \sum_n \exp(i \mathbf{Q} \cdot \mathbf{r}_{jkn}^\alpha).
\end{equation}
Here $\theta_{\alpha j}$ and $\mathbf{r}_{jkn}^\alpha$ are the domain fractions and atomic positions for the $\alpha$ terrace. 

Similar expressions apply to the reflectivity from the reconstructed layers on the $\beta$ terraces,
\begin{equation}
    r_\beta = \frac{r_f F_\beta}{m} \sum_{\ell = m_\alpha + 1}^{m} Y^\ell = \frac{r_f F_\beta}{m} \frac{Y(Y^m - Y^{m_\alpha})}{Y - 1},
     \label{eq:rb}
\end{equation}
\begin{equation}
    F_\beta = \sum_j \theta_{\beta j} \sum_k f_k \exp(-\sigma_k^2 Q^2) \sum_n \exp(i \mathbf{Q} \cdot \mathbf{r}_{jkn}^\beta).
\end{equation}

The total reflectivity amplitude is the sum of the complex amplitudes from the substrate and the reconstructed layers on the $\alpha$ and $\beta$ terraces,
\begin{equation}
    r_t = r_s + r_\alpha + r_\beta.
\end{equation}
The same expressions Eq.~(\ref{eq:dyn},\ref{eq:R}) given above relate the intensity reflectivity $R$ to $r_t$.

\begin{figure}
    \centering
    \includegraphics[width=1\linewidth]{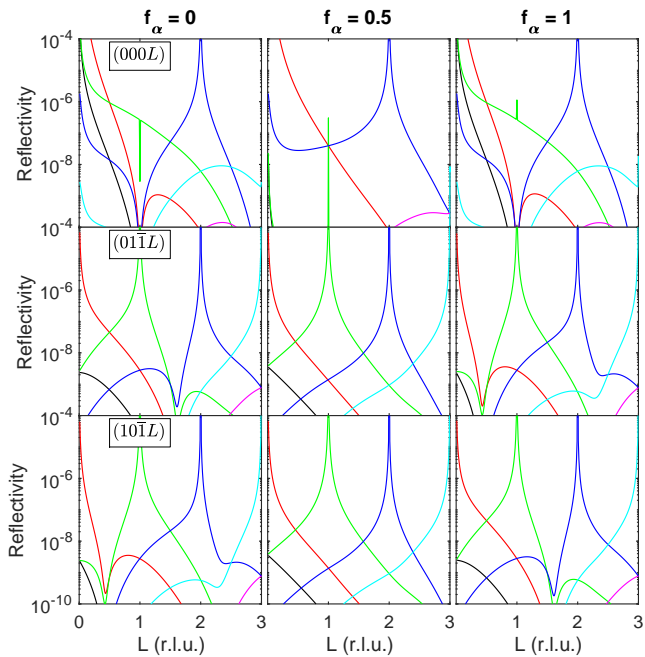} 
    \caption{Calculated reflectivities of CTRs for an $m = 100$ vicinal surface with the 3H(T1) reconstruction, $\sigma_R = 1$~\AA, and $\sigma_k = 0.11$~\AA. Top row: (0 0 0 L); middle row: $(0 1 \overline{1} L)$; bottom row: $(1 0 \overline{1} L)$. Black, red, green, blue, cyan, and magenta curves are for $L_0 = -1$ to $4$, respectively. Values of $f_\alpha$ for each column are given at the top. \label{fig:CTR_miscut_3HT1} }
\end{figure}

Figure \ref{fig:CTR_miscut_3HT1} shows the calculated reflectivity of the $(0 0 0 L_0)$, $(0 1 \overline{1} L_0)$, and $(1 0 \overline{1} L_0)$ CTRs for $L_0 = -1$ to $4$ for a miscut surface with three $f_\alpha$ values, 0.0, 0.5, and 1.0.
These calculations were done for a step period of $m = 100$, a surface with the 3H(T1) reconstruction with equal fractions $\theta_{\alpha j} = \theta_{\beta j} = 1/6$ of all domains on both terraces, a roughness of $\sigma_R = 1$~\AA, and a Debye-Waller length of $\sigma_k = 0.11$~\AA~ for all atoms.
The result is insensitive to 10\% changes in $m$.
While as in the case of an exactly oriented surface, the $(0 0 0 L_0)$ CTRs are identical for $f_\alpha = 0$ and $f_\alpha = 1$, they are very different for $f_\alpha = 0.5$, with the CTRs for even $L_0$ becoming stronger and the CTRs for odd $L_0$ becoming very weak.
The $(0 1 \overline{1} L_0)$ and $(1 0 \overline{1} L_0)$ CTRs have a more monotonic dependence on $f_\alpha$.
For $f_\alpha = 0$ and $f_\alpha = 1$, there are alternating stronger and weaker intensities between the Bragg peaks, with the alternation being opposite for $(0 1 \overline{1} L_0)$ and $(1 0 \overline{1} L_0)$.
For $f_\alpha = 0.5$, the intensities between the Bragg peaks are about the same, and there is no difference between the $(0 1 \overline{1} L_0)$ and $(1 0 \overline{1} L_0)$ CTRs.
The $(0 1 \overline{1} L_0)$ CTRs with $f_\alpha = X$ are identical to the $(1 0 \overline{1} L_0)$ CTRs with $f_\alpha = 1-X$, for any value $X$.
As with the exactly oriented surface, other reconstructions or HCP bulk structures show the same qualitative behavior.

\begin{figure}
\includegraphics[width=0.85\linewidth]{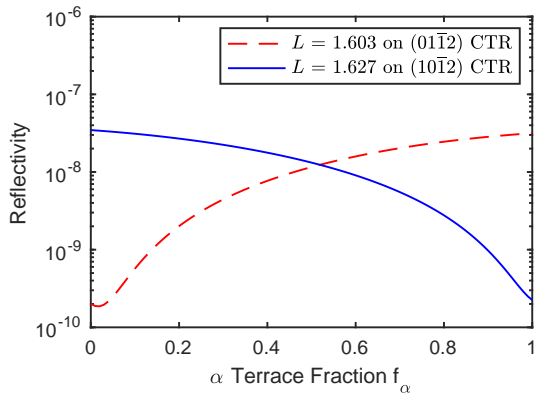}
\caption{Calculated reflectivity of selected CTRs as a function of terrace fraction $f_\alpha$, for fixed values of $L$ near $L = 1.6$. These curves use a fixed $\sigma_R = 0.9$~\AA~ and $\sigma_k = 0.11$~\AA. \label{fig:R_vs_f}}
\end{figure}

Figure \ref{fig:R_vs_f} shows calculations of the reflectivity as a function of $f_\alpha$ at positions near $L = 1.6$ on the $(0 1 \overline{1} 2)$ and $(1 0 \overline{1} 2)$ CTRs, for a surface with the 3H(T1) reconstruction.
Here we use a roughness of $\sigma_R = 0.9$~\AA~ to match the experimental fits described below.
The variation in reflectivity is almost monotonic in $f_\alpha$ at these positions.
These curves are used below to extract $f_\alpha(t)$ during dynamic transitions.

\begin{table*}
\caption{ \label{tab:table1} For each of four OMVPE conditions, we list the net growth rate $G$ in ML/s, where 1 ML $= c/2 = 2.6$~\AA, as well as values of $f_\alpha^{ss}$, $\sigma_R$, and $\chi^2$ from fits to reflectivity for each of 5 reconstructions.} 
\begin{ruledtabular}
\begin{tabular}{ c | c | c | c ||l | c | c | c | c | c }   
Growth & TEGa	flow & H$_2$ frac. &	Net growth & &3H(T1) & Ga(T4) & NH(H3)+ & NH(H3)+ & NH(H3)	\\
condition & ($\mu$mole & in &	rate & & & & H(T1) & NH$_2$(T1) &	\\
index & /min) & carrier &	(ML/s) & & & & & &	\\
\hline
& & & & $f_\alpha^{ss}$ & 0.111 &  0.144 & 0.098 & 0.106 & 0.095	\\
1&0.000 & 50\% & -0.0018 & $\sigma_R$(\AA) & 0.91 & 1.53 & 1.14 & 1.07 & 1.10 \\
& & & & $\chi^2$ & 106 & 130 & 187 & 200 & 167	\\
\hline
& & & & $f_\alpha^{ss}$ & 0.461 & 0.476 & 0.460 & 0.460 & 0.459\\
2 & 0.000 & 0\% & 0.0000 & $\sigma_R$(\AA)  & 1.13& 1.53 & 1.39 & 1.34 & 1.37 \\
& & & & $\chi^2$ & 57 & 81 & 76 &  67 & 99	\\
\hline
& & & & $f_\alpha^{ss}$ & 0.811 & 0.670 & 0.876 & 0.869 & 0.869	\\
3 & 0.033 & 50\% & 0.0109 & $\sigma_R$(\AA) & 1.03 & 1.77 & 1.44 & 1.40 & 1.40 \\
& & & & $\chi^2$ & 118 & 218 & 205 & 248 & 168	\\
\hline
 & & & & $f_\alpha^{ss}$ & 0.868 & 0.942 & 0.892 & 0.879 &  0.891	\\
4 & 0.033 & 0\% & 0.0127 & $\sigma_R$(\AA) & 0.57 & 1.28 & 1.09 & 1.03 & 1.05\\
& & & & $\chi^2$ & 80 & 112 & 174 & 220 & 135	\\
\end{tabular}
\end{ruledtabular}
\end{table*}

\section{Surface X-ray Scattering Measurements and Fits}

To characterize the behavior of $A$ and $B$ steps in GaN (0001) surfaces, we performed \textit{in situ} measurements of the CTRs during growth and evaporation in the OMVPE environment.
We used a chamber and goniometer at the Advanced Photon Source beamline 12ID-D, which was designed for \textit{in situ} surface X-ray scattering studies during growth \cite{2017_Ju_RSI88_035113}.
A micron-scale X-ray beam illuminated a small surface area having a uniform step azimuth.
To obtain sufficient signal, we used a wide-bandwidth ``pink'' beam setup similar to that described previously \cite{2018_Ju_JSyncRad25_1036,2019_Ju_NatPhys15_589}.
The beam incident on the sample had a typical intensity of $1.4 \times 10^{12}$ photons per second at $E = 25.75$ keV, in a spot size of $10 \times 10$ $\mu$m.
At the $2^\circ$ incidence angle, this illuminated an area of $10 \times 300$~$\mu$m.
X-ray scattering patterns were recorded using a photon counting area detector with a GaAs sensor having 512 $\times$ 512 pixels, 55 $\mu$m pixel size, located $1.1$ m from the sample (Amsterdam Scientific Instruments LynX 1800). 

Two types of measurements were performed.
We determined the steady-state terrace fractions $f_\alpha^{ss}$ under four different growth/evaporation conditions by scanning the detector along the $(0 1 \overline{1} L)$ and $(1 0 \overline{1} L)$ CTRs while continuously maintaining steady-state growth or evaporation.
We also observed the dynamics of the change in $f_\alpha$ by recording the intensity at a fixed detector position near $L = 1.6$ as a function of time before and after an abrupt change between conditions.

We studied four OMVPE conditions, summarized in Table~\ref{tab:table1}.
Under the conditions studied, deposition is transport limited, with the deposition rate proportional to the supply of the Ga precursor (triethylgallium, TEGa), with a large excess of the N precursor (NH$_3$) constantly supplied.
We investigated conditions of zero deposition (no supply of TEGa) as well as deposition at a TEGa supply of 0.033 $\mu$mole/min.
The NH$_3$ flow in both cases was 2.7 slpm or 0.12 mole/min, and the total pressure was 267 mbar. 
The V/III ratio during deposition was thus $3.6 \times 10^6$.
For both of these conditions, we studied two carrier gas compositions: 50\% H$_2$ + 50\% N$_2$, and 0\% H$_2$ + 100\% N$_2$.
The addition of H$_2$ to the carrier gas enhances evaporation of GaN, so that the net growth rate (deposition rate minus evaporation rate) is slightly lower; at zero deposition rate, the net growth rate is negative.
We determined the net growth rate for all four conditions as described in Appendix C.
These values are given in Table~\ref{tab:table1}.
Substrate temperatures were calibrated to within $\pm5$~K using laser interferometry from a standard sapphire substrate \cite{2017_Ju_RSI88_035113}.
While we used the same heater temperature for all conditions, the calibration indicates that the substrate temperature was slightly higher in 50\% H$_2$ (1080 K) than in 0\% H$_2$ (1073 K).

\begin{figure}
\includegraphics[width=0.75\linewidth]{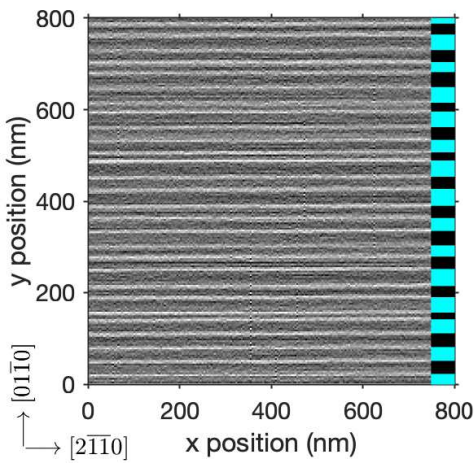}
\caption{\label{fig:AFM} AFM image of steps of height c/2 (2.6 \AA) on the GaN substrate used in X-ray measurements. To emphasize the positions of steps, we plot the amplitude error signal, which is proportional to the height gradient in the scan direction $(y)$. Image was obtained \textit{ex situ} at room $T$ after an anneal for 300s at 1118 K in zero-growth conditions (0\% H$_2$, 0 TEGa). Average fraction over a $2 \times 2$~$\mu$m area of ``even'' terraces (marked black at side) is 0.47. The average double-step spacing of $w = 57.3$~nm corresponds to an offcut of $\tan^{-1}(c/w) = 0.52^\circ$.} 
\end{figure}

\begin{figure}
\includegraphics[width=0.8\linewidth]{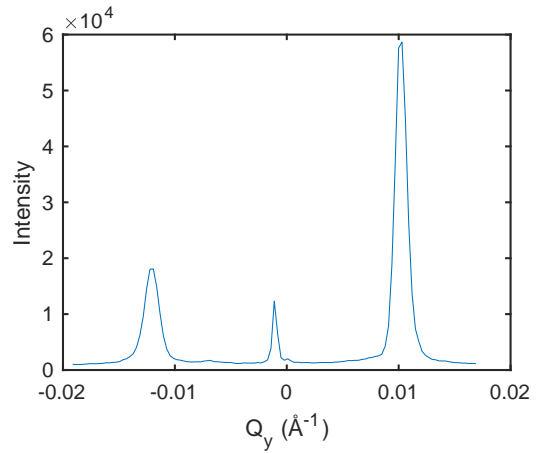}
\caption{\label{fig:00L} Profile of split CTRs showing offcut. Peaks correspond to the (0002), (0001), and (0000) CTRs at $L = 0.9$. Measured at $T = 1170$~K during growth at $0.053$~$\mu$mole/min TEGa, 50\% H$_2$. The splitting of the CTRs $\Delta Q_y = 0.0110$~\AA$^{-1}$ corresponds to an offcut of $\tan^{-1}[\Delta Q_y / (2 \pi / c)] = 0.52^\circ$.} 
\end{figure}

The substrate used was a GaN single crystal \footnote{GANKIBAN$^{TM}$ from SixPoint Materials, Inc., \url{spmaterials.com}.}. 
Figure~\ref{fig:AFM} shows its initial surface morphology determined by \textit{ex situ} atomic force microscopy (AFM) following an anneal at 1118 K for 300 s in zero-growth conditions (0\% H$_2$, 0 TEGa).
One can see straight steps almost perpendicular to $y$ over large areas.
An analysis of the step spacing shows a slight tendency towards pairing, with one of the two alternating terrace types having an area fraction of 0.47.
AFM is insensitive to whether this fraction corresponds to the $\alpha$ or $\beta$ terraces.
We also characterized the offcut by measuring the splitting of the CTRs.
Figure~\ref{fig:00L} shows a transverse cut through the CTRs in the $Q_y$ direction near (000L) at $L = 0.9$.
Both the AFM and X-ray measurements give a double-step spacing of $w = 573$~\AA~ corresponding to an offcut of $0.52^\circ$.
To relate the $\alpha$ terrace fraction to the behavior of $A$ and $B$ steps, it is critical to determine the sign of the step azimuth.
By making measurements as a function of $Q_z$, we verified that the peak at high $Q_y$ is the CTR coming from (0000), while the peak at low $Q_y$ is the (0002) CTR.
This confirms that the ``downstairs'' direction of the vicinal surface is in the $+y$ or $[0 1 \overline{1} 0]$ direction, as drawn in Fig.~\ref{fig:A_B}.
It is also useful to know the precise angle of the step azimuth with respect to the crystal planes, which determines the kink density and thus some kinetic coefficients.
X-ray measurements found this to be $5^\circ$ off of the $[0 1 \overline{1} 0]$ direction towards $[1 0 \overline{1} 0]$.
With this low-dislocation-density substrate and the low growth rates used, we did not observe the previously reported instability to step bunching during growth \cite{2000_Murty_PRB62_R10661}.

\begin{figure}
\includegraphics[width=0.9\linewidth]{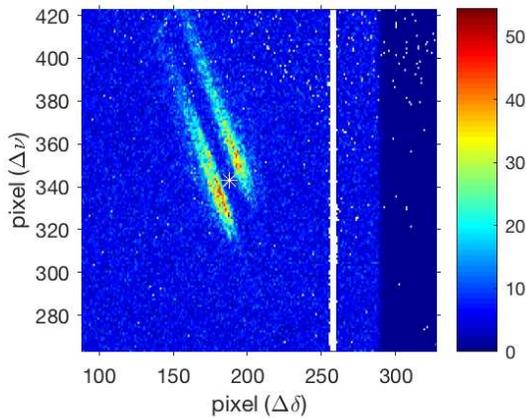}
\caption{\label{fig:CTR_angle_image} Detector image from $(0 1 \overline{1} L)$ scan at for condition 4 ($0.033 \mu$mol/min TEGa, 0\% H$_2$), showing intensity maxima from Ewald sphere cutting through CTRs from the $(0 1 \overline{1} 1)$ and $(0 1 \overline{1} 2)$ Bragg peaks. Position of central pixel, marked by cross, is $(0 1 \overline{1} L)$ with $L = 1.55$. Dark area on right is shadow of slits, white vertical line is gap in pixels between detector chips. White pixels are ignored due to excessive detector noise.}
\end{figure}

\begin{figure}
\includegraphics[width=0.9\linewidth]{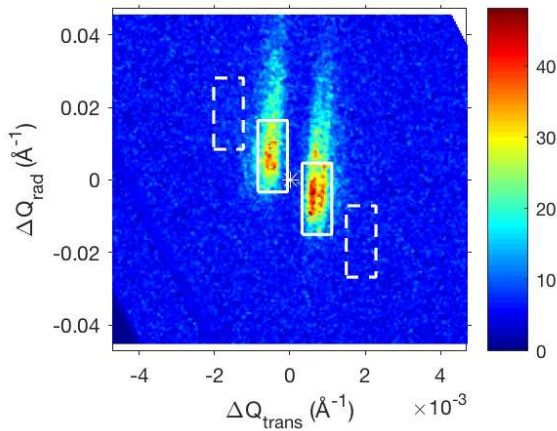}
\caption{\label{fig:CTR_qL_image} Cut through reciprocal space  at $L = Q_z c / 2 \pi = 1.55$, showing $(0 1 \overline{1} 1)$ and $(0 1 \overline{1} 2)$ CTRs. The in-plane $Q_x$ and $Q_y$ coordinates have been expressed as in-plane radial and transverse components $\Delta Q _{rad}$ and $\Delta Q_{trans}$ relative to the central pixel at position $(0 1 \overline{1} L)$. Rectangles give regions integrated to give CTR intensities and associated backgrounds.}
\end{figure}

\begin{figure}
\includegraphics[width=0.85\linewidth]{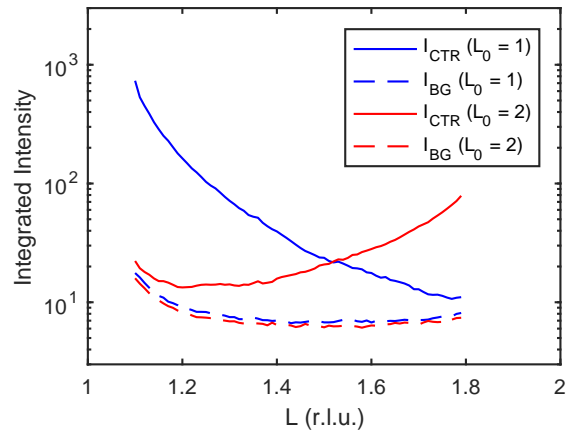}
\caption{\label{fig:CTR_backgrounds} Integrated total CTR and background intensities for the $(0 1 \overline{1} 1)$ and $(0 1 \overline{1} 2)$ CTRs, as a function of $L$ between $1.1$ and $1.8$, for condition 4 ($0.033$~ $\mu$mol/min TEGa, 0\% H$_2$).}
\end{figure}

\begin{figure*}
\includegraphics[width=1\linewidth]{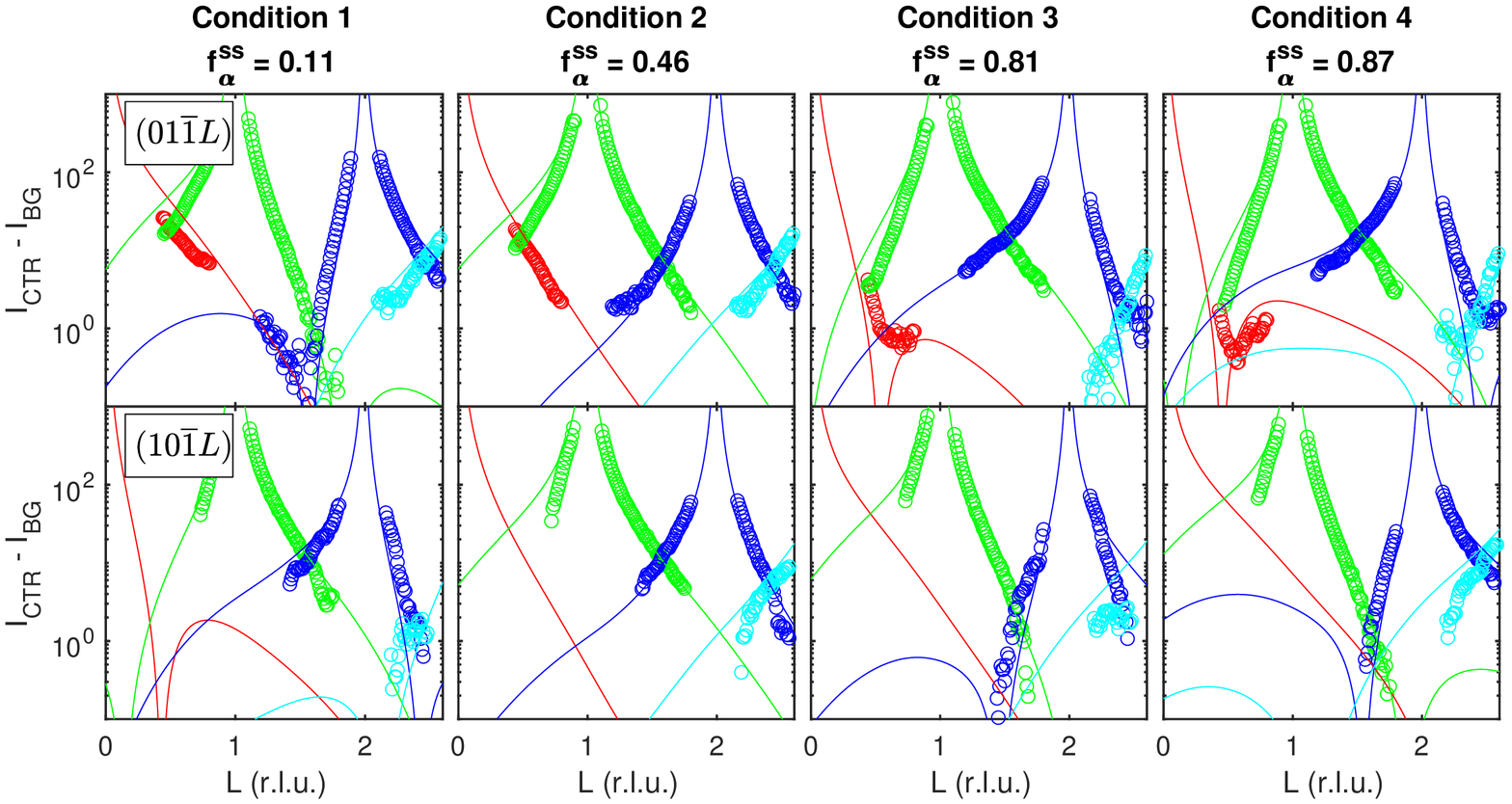}
\caption{Symbols show measured net intensities of the $(0 1 \overline{1} L_0)$ CTRs  (top row) and the $(1 0 \overline{1} L_0)$ CTRs (bottom row) for $L_0 = 0,1,2,3$ at each of four conditions. Curves show fits of all CTRs using the 3H(T1) reconstruction to obtain steady-state $\alpha$ terrace fraction $f_\alpha^{ss}$ at each condition.
\label{fig:fit_nonspecular_exp_3HT1}}
\end{figure*}

\begin{figure}
\includegraphics[width=0.9\linewidth]{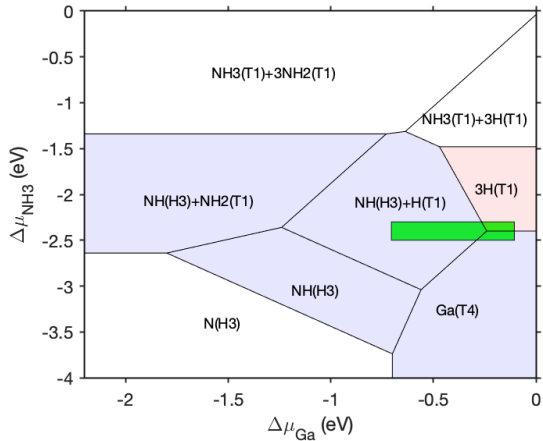}
\caption{\label{fig:recon_phase} Surface reconstruction phase diagram for GaN (0001), calculated in \cite{2012_Walkosz_PRB_85_033308}. Green rectangle shows estimated position of our experimental conditions, calculated in Appendix A. Five shaded reconstructions near these conditions were considered in fits shown in Table~\ref{tab:table1}.}
\end{figure}

\subsection{CTR measurements}

To process the X-ray data from the area detector, raw images were first corrected for detector flatfield, eliminating pixels with excessive noise, and the signal was normalized to the incident intensity. 
Figure~\ref{fig:CTR_angle_image} shows a typical corrected detector image, with streaks from the $(0 1 \overline{1} 1)$ and $(0 1 \overline{1} 2)$ CTRs.
Because of the $\sim 1\%$ energy bandwidth of the pink beam \cite{2019_Ju_NatPhys15_589}, the CTRs are broadened radially as well as being extended in the $Q_z$ direction.
To convert the images along an $L$ scan to reciprocal space, the $Q_x Q_y Q_z$ coordinates of each pixel in each image were first calculated.
The out-of-plane coordinate $Q_z$ or $L$ varies across each image, following the Ewald sphere.
The in-plane coordinates $Q_x$ and $Q_y$ were converted to in-plane radial and transverse components $\Delta Q_{rad}$ and $\Delta Q_{trans}$ relative to the central position.
The intensities and $L$ values of each image were interpolated onto a fixed grid of $\Delta Q_{rad}$ and $\Delta Q_{trans}$.
We then interpolate the sequence of intensities from the scan at each $\Delta Q_{rad}$ and $\Delta Q_{trans}$ onto a grid of fixed $L$ values.

Figure~\ref{fig:CTR_qL_image} shows a typical cut through reciprocal space at fixed $L$.
The peaks from the $(0 1 \overline{1} 1)$ and $(0 1 \overline{1} 2)$ CTRs are conveniently separated in $\Delta Q_{trans}$ because of the $5^\circ$ deviation of the step azimuth from $[0 1 \overline{1} 0]$; if the deviation was zero, the peaks would overlap at $\Delta Q_{trans} = 0$ because of the broadening in $\Delta Q_{rad}$.
Regions of $\Delta Q_{trans}$ and $\Delta Q_{rad}$ surrounding each CTR were defined to integrate the total intensity, with positions that vary with $L$ to follow the CTRs.
Likewise adjacent regions were defined to integrate an equivalent volume of background scattering.
Such regions are shown as rectangles in Fig.~\ref{fig:CTR_qL_image}.
Figure~\ref{fig:CTR_backgrounds} shows the mean total CTR intensities and backgrounds in these regions as a function of $L$ for the scan between $L = 1.1$ and $L = 1.8$ for condition 4.
The net CTR intensity was calculated by subtracting the background from the total for that CTR. 
We ran scans from $L = 0.4$ to $L = 0.9$, $L = 1.1$ to $L = 1.8$, and $L = 2.15$ to $L = 2.6$ on the $(0 1 \overline{1} L)$ and $(1 0 \overline{1} L)$ CTRs, skipping over the Bragg peaks to avoiding having the high intensity strike the detector.
The $L$ range covered on each CTR varied depending upon the region covered by the detector in reciprocal space during the scan.
Figure~\ref{fig:fit_nonspecular_exp_3HT1} shows the measured net CTR intensities as a function of $L$, for both the $(0 1 \overline{1} L)$ and $(1 0 \overline{1} L)$ CTRs and at all four conditions.
Only data points at which the total and background regions were fully captured on the detector without shadowing from the chamber window were kept.
This eliminated all of the data points for the $(1 0 \overline{1} 0)$ CTR.
The qualitative behavior agrees with that expected from a variation in $f_\alpha$ shown in Fig.~\ref{fig:CTR_miscut_3HT1}, with alternating higher and lower intensities between the Bragg peaks in some cases, and opposite behavior of the two CTRs. 

In order to determine whether exposure to the X-ray beam was affecting the OMVPE growth process, we periodically scanned the sample position while monitoring the CTR intensity.
For the conditions reported here, there was no indication that the spot which had been illuminated differed in any way from the neighboring regions.
During growth at higher temperatures (e.g. $1250$~K), we did observe local effects of the X-ray beam on the surface morphology.

\subsection{Fits to steady-state CTRs}

To obtain values of the steady-state terrace fraction $f_\alpha^{ss}$ for each of the four conditions, we fit the measured CTR intensities as a function of $L$ using the expressions developed in Section II above.
For each condition, the measurements of both the $(0 1 \overline{1} L)$ and $(1 0 \overline{1} L)$ CTRs were simultaneously fit.
In addition to a single value of $f_\alpha$, parameters varied in the fit included a surface roughness $\sigma_R$ and intensity scale factors for each CTR.
In the calculations we use $m = 100$, which produces negligible difference in $R$ compared with using the experimental value of $w/b = 103.4$.
Note that we allow $f_\alpha = m_\alpha/m$ to vary continuously, even though Eqs.~(\ref{eq:ra}) and (\ref{eq:rb}) were developed for integer $m_\alpha$.
We used equal fractions $\theta_{\alpha j} = \theta_{\beta j} = 1/6$ of all domains on both terraces, and fit to $\log(I)$ with equal weighting of all points.
Since no fractional-order diffraction peaks from long-range ordered reconstructions are observed, we expect that the domain structure has a short correlation length and all domains are present.

\begin{figure}
\includegraphics[width=0.8\linewidth]{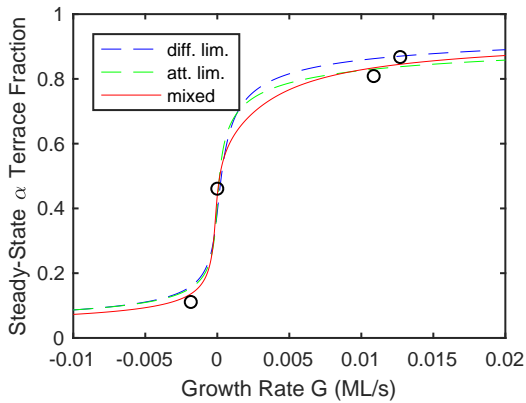}
\caption{Circles show experimental values of steady-state $\alpha$ terrace fraction as a function of growth rate obtained from fits to $(0 1 \overline{1} L)$ and $(1 0 \overline{1} L)$ CTRs using the 3H(T1) reconstruction, showing monotonic increase of $f_\alpha^{ss}$ with increasing $G$. Also shown are BCF model calculations described below. \label{fig:fss_exp}}
\end{figure}

\begin{figure}
\includegraphics[width=0.8\linewidth]{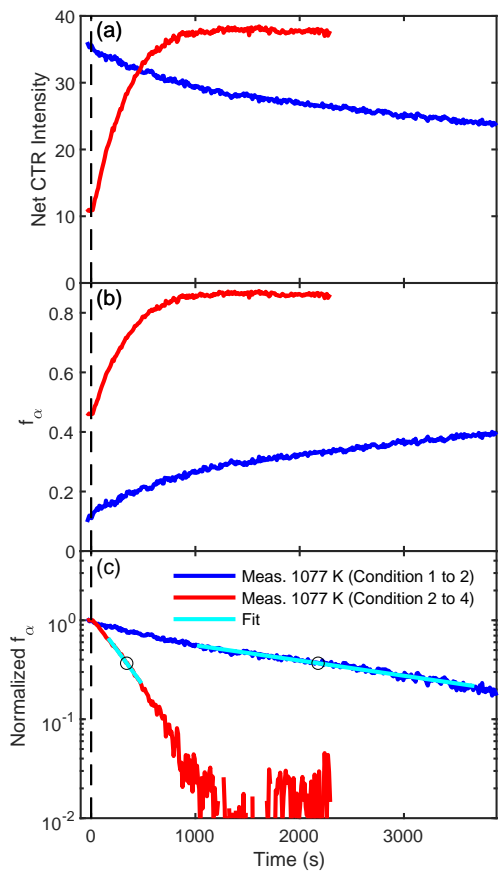}
\caption{Dynamics upon changing conditions. Blue curves: Evolution after changing from condition 1 to condition 2 at $t = 0$, determined from the $(1 0 \overline{1} 2)$ CTR at $L=1.627$. Red curves: Evolution after changing from condition 2 to condition 4 at $t = 0$, determined from the $(0 1 \overline{1} 2)$ CTR at $L=1.603$.
(a) Measured net CTR intensities.
(b) Evolution of terrace fraction $f_\alpha$ calculated from CTR intensities. 
(c) Normalized $f_\alpha$ change, plotted on a log scale, with a region fitted to a straight line to get the relaxation times $2177$ s and $341$ s as shown by circles.  \label{fig:relaxation_time}}
\end{figure}

We performed fits using different potential surface reconstructions.
Figure~\ref{fig:recon_phase} shows the calculated reconstruction phase diagram for the GaN (0001) surface in the OMVPE environment \cite{2012_Walkosz_PRB_85_033308}, as a function of Ga and NH$_3$ chemical potentials.
Based on the chemical potential values that correspond to our experimental conditions estimated in Appendix A, shown by the green rectangle, we considered the five reconstructions highlighted in Fig.~\ref{fig:recon_phase}.
(The estimate for $\Delta \mu_{Ga}$ has a large uncertainty because it depends on the nitrogen potential produced by decomposition of NH$_3$.)
Table~\ref{tab:table1} shows the values of $f_\alpha^{ss}$, $\sigma_R$, and the goodness-of-fit parameter $\chi^2$ from fits to reflectivity at four conditions for each of five reconstructions.
The qualitative results for the variation of $f_\alpha^{ss}$ with growth condition are independent of which reconstruction is assumed: $f_\alpha^{ss}$ increases monotonically as the net growth rate $G$ increases.
The 3H(T1) reconstruction gives the best fit (minimum $\chi^2$) of the five potential reconstructions, for all four conditions.
This is consistent with recent results on GaN (0001) reconstructions in the OMVPE environment, which found an even larger phase field for the 3H(T1) structure \cite{2019_Kempisty_PRB100_085304}.
Fig.~\ref{fig:fit_nonspecular_exp_3HT1} compares the fits with the 3H(T1) reconstruction to the measured CTR intensities.
Figure~\ref{fig:fss_exp} shows a plot of the resulting
 $f_\alpha^{ss}$ vs. net growth rate.
As we shall see below, the increase of $f_\alpha^{ss}$ with increasing growth rate indicates the nature of difference between the kinetics of adatom attachment at $A$ and $B$ steps: for GaN in the OMVPE environment, $A$ steps have faster kinetics.

\subsection{Dynamics of $f_\alpha$}

We also observed the dynamics of the change in $f_\alpha$ by recording the intensity at a fixed detector position as a function of time before and after an abrupt change between conditions, as shown in
Figure~\ref{fig:relaxation_time}(a). 
We chose positions near $L = 1.6$ where the reflectivity changes almost monotonically with $f_\alpha$, as shown in Fig.~\ref{fig:R_vs_f}.
It is thus straightforward to convert these intensity evolutions to variations in $f_\alpha$ by normalizing them to match the predicted change in reflectivity for the transition in $f_\alpha^{ss}$, and then inverting the $R(f_\alpha)$ relation to obtain $f_\alpha(t)$.
We assume the surface roughness is not a function of condition, and use the average value of $\sigma_R = 0.9$~\AA~ from the 3H(T1) fits to calculate $R(f_\alpha)$.
The resulting $f_\alpha(t)$ are shown in Fig.~\ref{fig:relaxation_time}(b).
To extract characteristic relaxation times for these transitions, we plot the normalized change in $f_\alpha$, i.e. $[f_\alpha - f_\alpha(t = \infty)]/[f_\alpha(t = 0) - f_\alpha(t = \infty)]$, on a log scale in Fig.~\ref{fig:relaxation_time}(c).
We fit the region indicated with a line and interpolated to obtain the $1/e$ decay point of these curves.

\section{Burton-Cabrera-Frank theory for vicinal c-plane surfaces}

To understand the behavior of the terrace fraction at steady-state and as a function of time after a change in growth rate, we have developed a model based on BCF theory for vicinal surfaces with a sequence of steps \cite{1999_Jeong_SurfSciRep34_171}.
This type of one-dimensional model considers adatom diffusion on terraces with boundary conditions at the steps defining the terrace edges, and has been used extensively to understand the step-bunching instability \cite{2020_Guin_PRL124_036101,2016_Li_ApplSurfSci371_242,2017_Bellmann_JCrystGrowth478_187,2007_Dufay_PRB75_241304,2003_Pierre-Louis_SurfSci529_114,2000_Pimpinelli_SurfSci445_L23}, step pairing \cite{2004_Pierre-Louis_PRL93_165901}, step width fluctuations \cite{2010_Patrone_PRE82_061601}, growth mode transitions \cite{2007_Ranguelov_PRB75_245419}, and competitive adsorption \cite{2019_Hanada_PhysRevMater3_103404}.
Typically, all steps in a sequence are assumed have identical properties.
In our case, we consider an alternating sequence of two types of terraces, $\alpha$ and $\beta$, and two types of steps, $A$ and $B$, with properties that can differ, as shown in Figs.~\ref{fig:A_B} and \ref{fig:BCF}.
Similar BCF models of alternating $A$ and $B$ steps have been considered previously \cite{2011_Zaluska-Kotur_JAP109_023515, 2010_Zaluska-Kotur_JNoncrystSolids356_1935,2006_Xie_PRB_085314}.
Here we include the effects of step transparency (also known as step permeability, the transmission of adatoms across steps) \cite{2003_Pierre-Louis_PRE68_021604,2003_Pierre-Louis_SurfSci529_114,2007_Ranguelov_PRB75_245419} and step-step repulsion \cite{1999_Jeong_SurfSciRep34_171,2010_Patrone_PRE82_061601}.

In this section we develop a quasi-steady-state expression for the dynamics of the terrace fraction $f_\alpha$, and give an exact solution using matrices.
Examples of the adatom distributions and $f_\alpha$ dynamics are shown.
Using further generally applicable assumptions, we develop a simplified analytical solution, and then consider cases of diffusion- or attachment-limited kinetics, and non-transparent or highly transparent steps.

\subsection{Exact quasi-steady-state solution}

\begin{figure}
\includegraphics[width=0.9\linewidth]{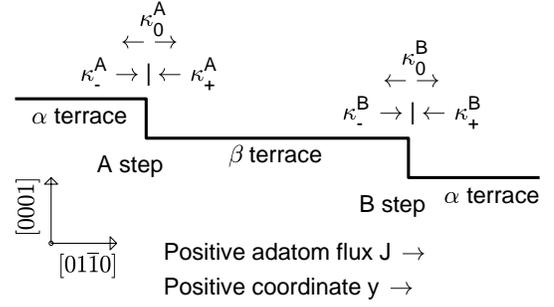}
\caption{Schematic of alternating terraces and steps of BCF model for HCP basal plane surfaces, showing kinetic coefficients for the $A$ and $B$ steps. \label{fig:BCF}}
\end{figure}

The continuity equation for the rate of change in the adatom density per unit area $\rho_i$ on terrace type $i = \alpha$ or $\beta$ is written as
\begin{equation}
    \frac{\partial \rho_i}{\partial t} = D \nabla^2 \rho_i - \frac{\rho_i}{\tau} + F,
    \label{eq:cont}
\end{equation}
where $D$ is the adatom diffusivity, $\tau$ is the adatom lifetime before evaporation, and $F$ is the deposition flux of adatoms per unit time and area. 
The four boundary conditions for the flux at the steps terminating each type of terrace can be written as
\begin{align}
    J_\alpha^+ &= - D\nabla \rho_\alpha^+
    = +\kappa_-^A (\rho_\alpha^+ - \rho_{eq}^A)
    + \kappa_0^A (\rho_\alpha^+ - \rho_\beta^-), \label{eq:bc1} \\
    J_\alpha^- &= - D\nabla \rho_\alpha^-
    = -\kappa_+^B (\rho_\alpha^- - \rho_{eq}^B)
    - \kappa_0^B (\rho_\alpha^- - \rho_\beta^+), \label{eq:bc2} \\
    J_\beta^+ &= - D\nabla \rho_\beta^+
    = +\kappa_-^B (\rho_\beta^+ - \rho_{eq}^B)
    + \kappa_0^B (\rho_\beta^+ - \rho_\alpha^-), \label{eq:bc3} \\
    J_\beta^- &= - D\nabla \rho_\beta^-
    = -\kappa_+^A (\rho_\beta^- - \rho_{eq}^A)
    - \kappa_0^A (\rho_\beta^- - \rho_\alpha^+), \label{eq:bc4}
\end{align}
where $J_i$ is the adatom flux on terrace $i$, $\kappa_+^j$ and $\kappa_-^j$ are the kinetic coefficients for adatom attachment at a step of type $j = A$ or $B$ from below or above, respectively,
$\kappa_0^j$ is the kinetic coefficient for transmission across the step,
$\rho_{eq}^j$ is the equilibrium adatom density at a step of type $j$, and the $+$ or $-$ superscripts on $J_i$, $\rho_i$, and $\nabla \rho_i$ indicate evaluation at the terrace boundaries $y = + w_i/2$ or $y = - w_i/2$, respectively, where $w_i$ is the width of the terraces of type $i$ and the spatial coordinate $y$ is taken to be zero in the center of each terrace. 
As shown in Fig.~\ref{fig:BCF}, the negative ``upstairs" boundary of a terrace of type $i = \alpha$ $(\beta)$ at $y = -w_i/2$ is a step of type $j = B$ $(A)$, while the positive ``downstairs" boundary at $y = +w_i/2$ is a step of type $j = A$ $(B)$, respectively.
A standard positive ES barrier is given by $\kappa_+^j > \kappa_-^j$.

The velocity $v_j$ of the $j$ type step can be obtained from the adatom fluxes arriving from each side, giving
\begin{align}
v_A &= \left ( J_\alpha^+ - J_\beta^- \right ) / \rho_0, \\
v_B &= \left ( J_\beta^+ - J_\alpha^- \right ) / \rho_0,
\end{align}
where $\rho_0$ is the density of lattice sites per unit area.
In the boundary conditions Eqs.~(\ref{eq:bc1}-\ref{eq:bc4}) we have neglected the ``advective'' terms $-\rho_i^\pm v_j$ due to the moving boundary \cite{2020_Guin_PRL124_036101}, under the assumption that the adatom coverages are small, $\rho_i << \rho_0$. 

We assume that the adatom density profiles $\rho_i(y)$ have reached a quasi-steady-state where we can set $\partial \rho_i / \partial t = 0$ in the continuity equation Eq.~(\ref{eq:cont}).
We still allow the terrace widths $w_i$ to evolve relatively slowly with time.
At quasi-steady-state, the general solution for the adatom densities satisfying Eq.~(\ref{eq:cont}) is
\begin{equation}
    \rho_i = F \tau + C_{1i} \cosh \left ( \frac{y}{\sqrt{D \tau}} \right ) + C_{2i} \sinh \left ( \frac{y}{\sqrt{D \tau}} \right ),
    \label{eq:qss}
\end{equation}
where $C_{1i}$ and $C_{2i}$ are coefficients to be determined from the boundary conditions for each terrace type $i = \alpha$ or $\beta$.
The gradient $\nabla \rho_i$ with respect to $y$ is then
\begin{equation}
    \nabla \rho_i = \frac{C_{1i}}{\sqrt{D \tau}} \sinh \left ( \frac{y}{\sqrt{D \tau}} \right ) 
    + \frac{C_{2i}}{\sqrt{D \tau}} \cosh \left ( \frac{y}{\sqrt{D \tau}} \right ).
    \label{eq:grad}
\end{equation}
If we define the coefficients
\begin{equation}
c_i \equiv \cosh \left ( \frac{w_i}{2\sqrt{D \tau}} \right ),
\end{equation}
\begin{equation}
s_i \equiv \sinh \left ( \frac{w_i}{2\sqrt{D \tau}} \right ),
\end{equation}
for terrace types $i = \alpha$ and $\beta$, and dimensionless step kinetic parameters
\begin{align}
p_j &\equiv (\tau/D)^{1/2} \, \, \kappa_+^j, \\
q_j &\equiv (\tau/D)^{1/2} \, \, \kappa_-^j, \\
r_j &\equiv (\tau/D)^{1/2} \, \, \kappa_0^j,
\end{align}
for step types $j = A$ and $B$,
then we can use the quasi-steady-state solution Eq.~(\ref{eq:qss},\ref{eq:grad}) to write the boundary conditions Eq.~(\ref{eq:bc1}-\ref{eq:bc4}) as
\begin{equation}
    \mathcal{M}\mathcal{C} = \mathcal{B}, \label{eq:MC}
\end{equation}
where $\mathcal{M}$ is a matrix given by
\begin{widetext}
\begin{equation}
\mathcal{M} =
\begin{bmatrix}
 + [ s_\alpha + (q_A + r_A) c_\alpha ] &
+ [ c_\alpha + (q_A + r_A) s_\alpha ] &
- r_A c_\beta & + r_A s_\beta \\
+ [ s_\alpha + (p_B + r_B) c_\alpha ] &
- [ c_\alpha + (p_B + r_B) s_\alpha ] &
- r_B c_\beta & - r_B s_\beta \\ 
- r_B c_\alpha & + r_B s_\alpha &
+ [ s_\beta + (q_B + r_B) c_\beta ] &
+ [ c_\beta + (q_B + r_B) s_\beta ] \\ 
- r_A c_\alpha & - r_A s_\alpha &
+ [ s_\beta + (p_A + r_A) c_\beta ] &
- [ c_\beta + (p_A + r_A) s_\beta ]
\end{bmatrix}
\label{eq:M}
\end{equation}
\end{widetext}
and the vectors $\mathcal{C}$ and $\mathcal{B}$ are given by
\begin{equation}
\mathcal{C} =
\begin{bmatrix}
C_{1\alpha} \\
C_{2\alpha} \\
C_{1\beta} \\
C_{2\beta}
\end{bmatrix},
\label{eq:C}
\end{equation}
\begin{equation}
\mathcal{B} =
\begin{bmatrix}
q_A (\rho_{eq}^A - F \tau) \\
p_B (\rho_{eq}^B - F \tau) \\ 
q_B (\rho_{eq}^B - F \tau) \\ 
p_A (\rho_{eq}^A - F \tau)
\end{bmatrix}.
\label{eq:B}
\end{equation}
The solution for the values of the four coefficients $C_{1i}$ and $C_{2i}$ of Eq.~(\ref{eq:qss}) is given by
\begin{equation}
    \mathcal{C} = \mathcal{M}^{-1} \mathcal{B},
\label{eq:MB}
\end{equation}
where $\mathcal{M}^{-1}$ is the inverse of $\mathcal{M}$.

The quasi-steady-state step velocities can then be evaluated from expressions obtained using Eqs.~(\ref{eq:bc1}-\ref{eq:grad}),
\begin{align}
v_A &= - \sqrt{\frac{D}{\tau}} \left ( 
\frac{s_\alpha C_{1\alpha} + c_\alpha C_{2\alpha} 
+ s_\beta C_{1\beta} - c_\beta C_{2\beta}}
{\rho_0} \right ), \label{eq:va} \\
v_B &= - \sqrt{\frac{D}{\tau}} \left ( 
\frac{s_\alpha C_{1\alpha} - c_\alpha C_{2\alpha} 
+ s_\beta C_{1\beta} + c_\beta C_{2\beta}}
{\rho_0} \right ). \label{eq:vb}
\end{align}

The final relationships needed are those between the equilibrium adatom densities at the steps $\rho_{eq}^j$ and the terrace widths.
These relationships reflect an effective repulsion between the steps owing to entropic and strain effects \cite{1999_Jeong_SurfSciRep34_171,2010_Patrone_PRE82_061601}.
In our case, with two different types of steps, we use the relations
\begin{equation}
    \rho_{eq}^j = \rho_{eq}^0 \exp(\mu_j/kT),
\end{equation}
where $\rho_{eq}^0$ is the equilibrium adatom density at zero growth rate, and the chemical potentials $\mu_j$ for the $j = A$ and $B$ steps are
\begin{equation}
    \frac{\mu_A}{kT} = -\frac{\mu_B}{kT} = \mathbf{M} = \left ( \frac{\ell_\beta}{w_\beta} \right )^3 - \left ( \frac{\ell_\alpha}{w_\alpha} \right)^3.
\end{equation}
Here the $\ell_i$ are two step repulsion lengths, that can differ for the two types of terraces.

We consider the overall vicinal angle of the surface to fix the sum $w$ of the widths of $\alpha$ and $\beta$ terraces, so that the widths can be expressed as $w_i = f_i w$, where there is one independent terrace fraction $f_\alpha$, and the other is given by $f_\beta = 1 - f_\alpha$.
In this case we can express the step chemical potentials as
\begin{equation}
     \mathbf{M}(f_\alpha) = \left ( \frac{\ell}{w} \right )^3 \left [ \left ( \frac{1 - f_\alpha^0}{1 - f_\alpha} \right )^3 - \left ( \frac{f_\alpha^0}{f_\alpha} \right)^3 \right ],
\end{equation}
where the coefficients $\ell$ and $f_\alpha^0$ are related to the $\ell_i$ by
\begin{align}
    \ell_\alpha &= f_\alpha^0 \ell, \\
    \ell_\beta &= (1 - f_\alpha^0) \ell.
\end{align}
where $f_\alpha^0 $ is the terrace fraction at zero growth rate.

The net growth rate $G$ in monolayers per second is proportional to the sum of the step velocities,
\begin{equation}
 G = \frac{v_A + v_B}{w} = - \sqrt{\frac{D}{\tau}} \left ( 
\frac{2 s_\alpha C_{1\alpha} 
+ 2 s_\beta C_{1\beta}}
{w \rho_0} \right ).
\end{equation}
The rate of change of the $\alpha$ terrace fraction $f_\alpha$ is proportional to the step velocity difference,
\begin{equation}
 \frac{df_\alpha}{dt} = \frac{v_A - v_B}{w} = \sqrt{\frac{D}{\tau}} \left ( 
\frac{2 c_\beta C_{2\beta}
- 2 c_\alpha C_{2\alpha}}
{w \rho_0} \right ).
\label{eq:dfdt}
\end{equation}
This equation can be integrated to solve for the evolution of $f_\alpha(t)$ at quasi-steady-state.
To obtain the full steady-state value of $f_\alpha$, the $A$ and $B$ step velocities must be equal and stable against fluctuations,
\begin{equation}
    \frac{df_\alpha}{dt} = 0,
    \label{eq:fullss}
\end{equation}
\begin{equation}
    \frac{\partial (df_\alpha/dt)}{\partial f_\alpha} < 0.
\end{equation}
When the net growth rate is zero and the terrace fraction has reached it full steady-state value, the step velocities are both zero, the diffusion fluxes are zero, the adatom densities are constant at a value $\rho_\alpha = \rho_\beta = \rho_{eq}^A = \rho_{eq}^B = \rho_{eq}^0$, and $\mu_A = -\mu_B = 0$. 
One can see that the parameter $f_\alpha^0$ is the full steady-state value of $f_\alpha$ at zero growth rate.

\begin{figure}
\includegraphics[width=0.78\linewidth]{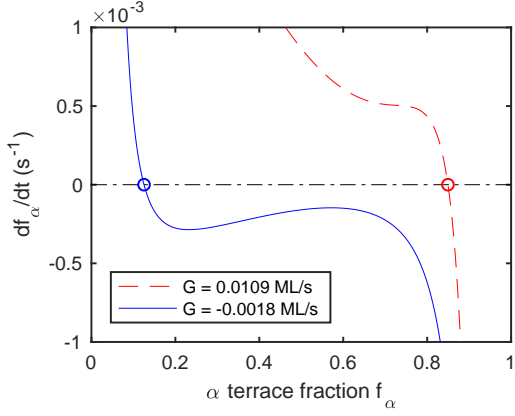}
\caption{Rate of change of the terrace fraction $df_\alpha/dt$ as a function of terrace fraction $f_\alpha$, calculated from Eq.~(\ref{eq:dfdt}) with parameter values given in Table~\ref{tab:table2}. The steady-state values of $f_\alpha$ are marked with a circle.
\label{fig:dfdt1}}
\end{figure}

\begin{figure}
\includegraphics[width=0.78\linewidth]{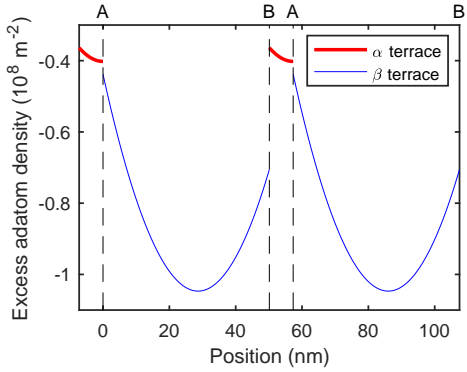}
\caption{Excess adatom density $\rho_i - \rho_{eq}^0$ on a sequence of $\alpha$ and $\beta$ terraces corresponding to the steady-state solution, calculated with parameter values given in Table~\ref{tab:table2}, for $F = 0$, $G = -0.00184$~ML/s. \label{fig:rhovy1}}
\end{figure}

\subsection{Calculation of steady-state and dynamics}

Here we show some examples calculated from the BCF theory. Figure~\ref{fig:dfdt1} shows the quasi-steady-state rate of change of the terrace fraction $df_\alpha/dt$ as a function of terrace fraction $f_\alpha$, calculated from Eq.~(\ref{eq:dfdt}) with parameter values given in Table~\ref{tab:table2}.
One curve is for a situation with no deposition flux, $F = 0$, where evaporation causes the net growth rate to be negative, $G = -0.00184$~ML/s, while the other is for a deposition flux of $F = 1.43\times10^{17}$~m$^{-2}$s$^{-1}$, giving a positive net growth rate of $G = 0.0109$~ML/s.
The steady-state values of $f_\alpha$ where $df_\alpha/dt = 0$ are marked.
For these parameters there is only a single steady-state solution for each curve, but from the non-monotonic shapes of the curves, one can see that two stable steady-state solutions can occur.

\begin{figure}
\includegraphics[width=0.78\linewidth]{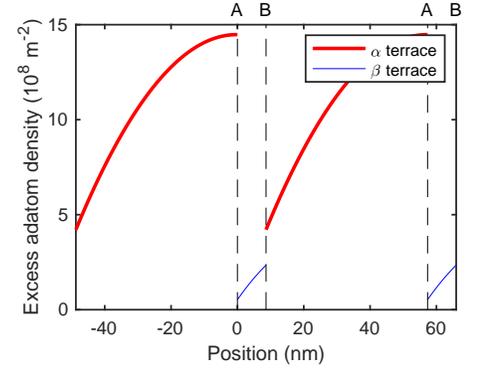}
\caption{Excess adatom density $\rho_i - \rho_{eq}^0$ on a sequence of $\alpha$ and $\beta$ terraces corresponding to the steady-state solution, calculated with parameter values given in Table~\ref{tab:table2}, for $F = 1.43 \times 10^{17}$~m$^{-2}$s$^{-1}$, $G = 0.0109$~ML/s. \label{fig:rhovy2}}
\end{figure}

\begin{figure}
\includegraphics[width=0.78\linewidth]{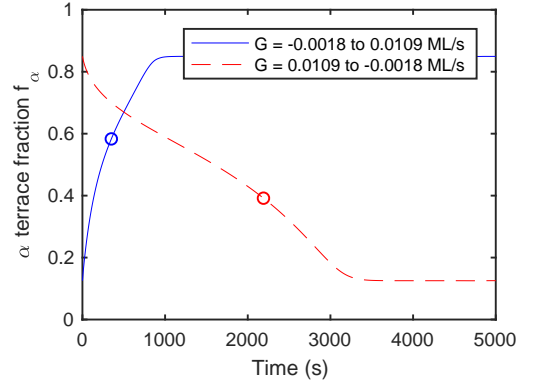}
\caption{Time dependence of $f\alpha$ obtained by integrating the quasi-steady-state result, Eq.~(\ref{eq:dfdt}), following changes between $F=0$ and $F=1.43 \times 10^{17}$~m$^{-2}$s$^{-1}$. The circles show examples of $1/e$ relaxation times. \label{fig:fvst1}}
\end{figure}

\begin{table} 
\caption{ \label{tab:table2} Parameter values used in BCF theory calculations shown in Figs.~\ref{fig:dfdt1} - \ref{fig:fvst1}, from fits and estimates given below.}
\begin{ruledtabular}
\begin{tabular}{ l  l }   
$w = 5.73 \times 10^{-8}$ m & $\rho_0 = 1.13 \times 10^{19}$ m$^{-2}$ \\
$\ell = 8.8 \times 10^{-10}$ m & $\rho_{eq}^0 = 3.5 \times 10^{11}$ m$^{-2}$ 	\\
$\tau = 1.7 \times 10^{-5}$ s & $D = 1.4 \times 10^{-7}$ m$^2$/s \\
$\kappa_+^A = 1.0 \times 10^4$ m/s & $\kappa_+^B = 7.4 \times 10^0$ m/s  \\
$\kappa_-^A = 1.0 \times 10^{-2}$ m/s & $\kappa_-^B = 1.0 \times 10^{-2}$ m/s  \\
$\kappa_0^A = 1.0 \times 10^{-2}$ m/s & $\kappa_0^B = 1.3 \times 10^1$ m/s  \\
$f_\alpha^0 = 0.44$ & $F = 0$ or $1.43 \times 10^{17}$ m$^{-2}$s$^{-1}$  \\
\end{tabular}
\end{ruledtabular}
\end{table}

Figures~\ref{fig:rhovy1} and \ref{fig:rhovy2} show the distribution of adatom density on a sequence of $\alpha$ and $\beta$ terraces at steady-state.
Since the deviations from $\rho_{eq}^0$ are very small, these are shown as the excess density $\rho_i - \rho_{eq}^0$.
In Fig.~\ref{fig:rhovy1}, where $G$ is negative (i.e. evaporation is faster than deposition), the densities tend to go through minima on each terrace, while in Fig.~\ref{fig:rhovy2}, $G$ is positive (i.e. deposition is faster than evaporation), the densities tend to go through maxima.
The low values of $\kappa_-^A$ and $\kappa_-^B$ used imply large ES barriers at the downhill (positive $y$) edges of the terraces, moving the maximum or minimum to that side.
The value of $\kappa_0^B$ gives significant transport across the $B$ step, reducing difference in adatom densities across the step. 

Figure~\ref{fig:fvst1} shows the calculated time dependence of $f_\alpha$ obtained by integrating the quasi-steady-state result, Eq.~(\ref{eq:dfdt}), for changes between the two conditions of $F = 0$ and $F = 1.43 \times 10^{17}$~m$^{-2}$s$^{-1}$.
While the predicted shapes are not simple exponentials in these cases, for fitting to experiments we nonetheless characterize the model dynamics using the time to reach the $1/e$ fraction of the change in steady-state $f_\alpha^{ss}$.

\subsection{Analytical solution for non-transparent steps}

Because all four boundary conditions implied by Eq.~(\ref{eq:MC}) involve terms in all four coefficients $C_{1i}$ and $C_{2i}$, the explicit analytical solution of Eq.~(\ref{eq:MB}) for the coefficients gives very elaborate expressions. 
In the case of non-transparent steps, with $r_A = r_B = 0$, half of the elements of $\mathcal{M}$ drop out and the boundary conditions split into two sets of two equations, each involving only two coefficients. 
In this case the analytical solutions are
\begin{widetext}
\begin{align}
C_{1\alpha} &= \frac{-F \tau [2 p_B q_A s_\alpha + (p_B + q_A)c_\alpha] 
+ (\rho_{eq}^A + \rho_{eq}^B) p_B q_A s_\alpha
+ (q_A \rho_{eq}^A + p_B \rho_{eq}^B) c_\alpha}
{(p_B + q_A)(s_\alpha^2 + c_\alpha^2)
+ 2(1 + p_B q_A) s_\alpha c_\alpha}, \label{eq:Ci1} \\
C_{2\alpha} &= \frac{F \tau (p_B - q_A) s_\alpha
+ (\rho_{eq}^A - \rho_{eq}^B) p_B q_A c_\alpha
+ (q_A \rho_{eq}^A - p_B \rho_{eq}^B) s_\alpha}
{(p_B + q_A)(s_\alpha^2 + c_\alpha^2)
+ 2(1 + p_B q_A) s_\alpha c_\alpha}, \label{eq:Ci2} \\
C_{1\beta} &= \frac{-F \tau [2 p_A q_B s_\beta + (p_A + q_B)c_\beta] 
+ (\rho_{eq}^B + \rho_{eq}^A) p_A q_B s_\beta
+ (q_B \rho_{eq}^B + p_A \rho_{eq}^A) c_\beta}
{(p_A + q_B)(s_\beta^2 + c_\beta^2)
+ 2(1 + p_A q_B) s_\beta c_\beta}, \label{eq:Ci3} \\
C_{2\beta} &= \frac{F \tau (p_A - q_B) s_\beta
+ (\rho_{eq}^B - \rho_{eq}^A) p_A q_B c_\beta
+ (q_B \rho_{eq}^B - p_A \rho_{eq}^A) s_\beta}
{(p_A + q_B)(s_\beta^2 + c_\beta^2)
+ 2(1 + p_A q_B) s_\beta c_\beta}. \label{eq:Ci4} 
\end{align}
\end{widetext}

\subsection{Analytical solution for transparent steps}

To obtain an analytical solution of Eq.~(\ref{eq:MB}) including the effects of step transparency, we can work with an alternative, equivalent formulation of the boundary conditions \cite{2003_Pierre-Louis_PRE68_021604}

\begin{align}
    J_\alpha^+ &= - D\nabla \rho_\alpha^+
    = +\kapt_-^A (\rho_\alpha^+ - \rhot_{eq}^A), \label{eq:bca1} \\
    J_\alpha^- &= - D\nabla \rho_\alpha^-
    = -\kapt_+^B (\rho_\alpha^- - \rhot_{eq}^B), \label{eq:bca2} \\
    J_\beta^+ &= - D\nabla \rho_\beta^+
    = +\kapt_-^B (\rho_\beta^+ - \rhot_{eq}^B), \label{eq:bca3} \\
    J_\beta^- &= - D\nabla \rho_\beta^-
    = -\kapt_+^A (\rho_\beta^- - \rhot_{eq}^A), \label{eq:bca4}
\end{align}
where the quantities with tildes are defined as
\begin{align}
\kapt_+^j &\equiv \frac{\kapsq^j}{\kappa_-^j}, \label{eq:kt1} \\
\kapt_-^j &\equiv \frac{\kapsq^j}{\kappa_+^j}, \label{eq:kt2} \\
\rhot_{eq}^j &\equiv \rho_{eq}^j + \frac{v_j \rho_0 \kappa_0^j}{\kapsq^j}, \label{eq:kt3} \\
\kapsq^j &\equiv \kappa_+^j \kappa_-^j + \kappa_+^j \kappa_0^j + \kappa_-^j \kappa_0^j. \label{eq:kt4}
\end{align}
Note that in Eq.~(\ref{eq:kt3}) the effective equilibrium adatom density $\rhot_{eq}^j$ at a step of type $j$ depends on the step velocity $v_j$.
The boundary conditions can be written as
\begin{equation}
    \tilde{\mathcal{M}} \mathcal{C} = \tilde{\mathcal{B}}, \label{eq:MCt}
\end{equation}
where $\tilde{\mathcal{M}}$ and $\tilde{\mathcal{B}}$ are given by
\begin{align}
&\tilde{\mathcal{M}} = \nonumber \\
&\begin{bmatrix}
s_\alpha + \qt_A c_\alpha &
c_\alpha + \qt_A s_\alpha &
0 & 0 \\
s_\alpha + \pt_B c_\alpha &
- c_\alpha - \pt_B s_\alpha &
0 & 0 \\ 
0 & 0 &
s_\beta + \qt_B c_\beta &
c_\beta + \qt_B s_\beta \\ 
0 & 0 &
s_\beta + \pt_A c_\beta &
- c_\beta - \pt_A s_\beta
\end{bmatrix},
\label{eq:Mt}
\end{align}
\begin{equation}
\tilde{\mathcal{B}} =
\begin{bmatrix}
\qt_A (\rhot_{eq}^A - F \tau) \\
\pt_B (\rhot_{eq}^B - F \tau) \\ 
\qt_B (\rhot_{eq}^B - F \tau) \\ 
\pt_A (\rhot_{eq}^A - F \tau)
\end{bmatrix},
\label{eq:Bt}
\end{equation}
using new dimensionless step kinetic parameters
\begin{align}
\pt_j &\equiv \sqrt{\frac{\tau}{D}} \kapt_+^j = \frac{p_j q_j + p_j r_j + q_j r_j}{q_j}, \\
\qt_j &\equiv \sqrt{\frac{\tau}{D}} \kapt_-^j = \frac{p_j q_j + p_j r_j + q_j r_j}{p_j},
\label{eq:ptqt}
\end{align}
for step types $j = A$ and $B$.
As in the case of non-transparent steps, these boundary conditions consist of two sets of two equations, each involving only two coefficients, $C_{1i}$ and $C_{2i}$ with $i = \alpha$ or $\beta$. 
The solutions are the same as Eqs.~(\ref{eq:Ci1}-\ref{eq:Ci4}), with $p_j$, $q_j$, and $\rho_{eq}^j$ replaced by $\pt_j$, $\qt_j$, and $\rhot_{eq}^j$, respectively.
Unfortunately, since the $\rhot_{eq}^j$ that appear in the $C_{1i}$ and $C_{2i}$ depend upon the step velocities $v_j$, which in turn depend upon the $C_{1i}$ and $C_{2i}$ via Eqs.~(\ref{eq:va}-\ref{eq:vb}), this still does not provide an explicit solution for the $C_{1i}$ and $C_{2i}$.

\subsection{Simplified analytical solution}

It is very useful to consider some generally applicable limits which simplify the analytical solution, allowing the steady-state terrace fraction and its dynamics to to be expressed in terms of the net growth rate.
We start with Eqs.~(\ref{eq:Ci1}-\ref{eq:Ci4}), with $p_j$, $q_j$, and $\rho_{eq}^j$ replaced by $\pt_j$, $\qt_j$, and $\rhot_{eq}^j$, respectively.
In the limit where the diffusion length within an adatom lifetime is much larger than the terrace widths, $\sqrt{D \tau} >> w$, the coefficients $c_i$ can be set equal to unity, and the coefficients $s_i$ are small quantities given by $s_i = w_i/(2 \sqrt{D \tau})$.
In the limit $\ell_i << w_i$, the adatom densities $\rho_i$ do not differ much from $\rho_{eq}^0$, and thus the adatom evaporation flux is relatively uniform at $\rho_{eq}^0/\tau$.
Assuming the second term in Eq.~(\ref{eq:kt3}) is small,
we can replace $\rhot_{eq}^A$ and $\rhot_{eq}^B$ by $\rho_{eq}^0$, except in the difference $(\rhot_{eq}^A-\rhot_{eq}^B)$.
We check the self-consistency of this assumption below.
If we also assume that the attachment parameters are generally greater than unity, so that $\pt_A \qt_B >> 1$, $\pt_B \qt_A >> 1$, the formulas for $C_{1i}$ simplify to be 
\begin{equation}
C_{1\alpha} \approx C_{1\beta} \approx \rho_{eq}^0 - F \tau.  \label{eq:Ca1}
\end{equation}
The net growth rate is then simply given by
\begin{equation}
    G \approx \frac{F - \rho_{eq}^0 / \tau}{\rho_0},  \label{eq:Ga}
\end{equation}
which is the difference between the deposition flux $F$ and a uniform evaporation flux $\rho_{eq}^0 / \tau$, converted to ML/s using $\rho_0$.
We can write the expressions for the $C_{2i}$ as
\begin{align}
C_{2\alpha} &\approx \frac{\sqrt{D \tau}}{w} \big [ R_\alpha (\rhot_{eq}^A - \rhot_{eq}^B) + S_\alpha \rho_0 G \big ], \label{eq:Ca2a} \\
C_{2\beta} &\approx \frac{\sqrt{D \tau}}{w} \big [ R_\beta (\rhot_{eq}^B - \rhot_{eq}^A) + S_\beta \rho_0 G \big ], \label{eq:Ca3a}
\end{align}
where each contains a term that is proportional to the net growth rate $G$.
The coefficients are give by
\begin{align}
R_\alpha &\equiv \frac{w}{D} \left ( \frac{\kappa_+^A}{\kapsq^A} + \frac{\kappa_-^B}{\kapsq^B} + \frac{w f_\alpha}{D} \right )^{-1},  \label{eq:Ra} \\
R_\beta &\equiv  \frac{w}{D} \left ( \frac{\kappa_+^B}{\kapsq^B} + \frac{\kappa_-^A}{\kapsq^A} + \frac{w (1 - f_\alpha)}{D} \right )^{-1},  \label{eq:Rb} \\
S_\alpha &\equiv \frac{R_\alpha w f_\alpha}{2} \left ( \frac{\kappa_+^A}{\kapsq^A} - \frac{\kappa_-^B}{\kapsq^B} \right ),  \label{eq:Sa} \\
S_\beta &\equiv  \frac{R_\beta w (1 - f_\alpha)}{2} \left ( \frac{\kappa_+^B}{\kapsq^B} - \frac{\kappa_-^A}{\kapsq^A} \right ),  \label{eq:Sb}
\end{align}
where the $R_i$ are positive and dimensionless and the $S_i$ have dimensions of time.
The step velocities of Eqs.~(\ref{eq:va}-\ref{eq:vb}) become
\begin{align}
    v_A &= \frac{w G}{2} + \frac{D}{\rho_0 w} \big [ (R_\alpha + R_\beta) (\rhot_{eq}^B - \rhot_{eq}^A) + (S_\beta - S_\alpha) \rho_0 G \big ],  \label{eq:vas}\\
    v_B &= \frac{w G}{2} + \frac{D}{\rho_0 w} \big [ (R_\alpha + R_\beta) (\rhot_{eq}^A - \rhot_{eq}^B) + (S_\alpha - S_\beta) \rho_0 G \big ]. \label{eq:vbs}
\end{align}
The difference of the effective equilibrium step adatom densities also contains a term that is proportional to $G$,
\begin{equation}
    \rhot_{eq}^A - \rhot_{eq}^B = \frac{2\rho_{eq}^0 \mathbf{M} + \rho_0 G \big [ S_0 + R_0 (S_\beta - S_\alpha) \big ]}{1 + R_0 (R_\alpha + R_\beta)},
\end{equation}
where the new coefficients are given by
\begin{align}
    R_0 &\equiv \frac{D}{w} \left ( \frac{\kappa_0^A}{\kapsq^A} + \frac{\kappa_0^B}{\kapsq^B} \right ), \label{eq:R0} \\
    S_0 &\equiv \frac{w}{2} \left ( \frac{\kappa_0^A}{\kapsq^A} - \frac{\kappa_0^B}{\kapsq^B} \right ).  \label{eq:S0}
\end{align}
The rate of change of $f_\alpha$ becomes
\begin{equation}
    \frac{df_\alpha}{dt} = \mathbf{K}^{dyn}(f_\alpha) \left ( \frac{G}{\mathbf{K}^{ss}(f_\alpha)} - \frac{4 \mathbf{M}(f_\alpha) \rho_{eq}^0}{w \rho_0} \right ), \label{eq:dfdta}
\end{equation}
where we have introduced the combined kinetic coefficient functions $\mathbf{K}^{ss}(f_\alpha)$ and $\mathbf{K}^{dyn}(f_\alpha)$, defined by
\begin{align}
    \mathbf{K}^{ss}(f_\alpha) &\equiv \frac{w}{2 \big [ -S_0 + (S_\beta - S_\alpha)/(R_\alpha + R_\beta) \big ]},
    \label{eq:Dk} \\
    \mathbf{K}^{dyn}(f_\alpha) &\equiv \frac{D (R_\alpha + R_\beta)}{w [ 1 + R_0 (R_\alpha + R_\beta)]}.
    \label{eq:KD}
\end{align}
These functions have the same dimensions as the individual $\kappa_x^j$ coefficients (length/time). 
$\mathbf{K}^{dyn}(f_\alpha)$ is always positive; $\mathbf{K}^{ss}(f_\alpha)$ depends on the differences in the $\kappa_x^j$, such that in the limit where all $\kappa_x^j$ are equal, $\mathbf{K}^{ss} \rightarrow \infty$.
In this case the influence of $G$ on $f_\alpha$ becomes negligible, and the steady-state $\alpha$ terrace fraction is always $f_\alpha^{ss} = f_\alpha^0$ (i.e. the value where $\mathbf{M} = 0$), independent of $G$. 

The general equation to obtain the full steady state is
\begin{equation}
    G^{ss}(f_\alpha) = \frac{4 \, \mathbf{K}^{ss}(f_\alpha) \, \mathbf{M}(f_\alpha) \rho_{eq}^0}{w \rho_0}.
    \label{eq:fSSG}
\end{equation}
This equation for $G^{ss}(f_\alpha)$ can be inverted to obtain a master curve for the steady-state value $f_\alpha^{ss}$ as a function of $G$.
For both the dynamics Eq.~(\ref{eq:dfdta}) and the steady-state Eq.~(\ref{eq:fSSG}),
the six step attachment parameters enter through the six combinations in the coefficients $R_i$, $S_i$, $R_0$, and $S_0$. 
The only dependence on $\tau$ and $F$ is through their combination into $G$, Eq.~(\ref{eq:Ga}).

The curve $G^{ss}(f_\alpha)$ always passes through $G = 0$ at $f_\alpha = f_\alpha^0$, since $\mathbf{M}$ is zero there.
The slope of the curve at $f_\alpha = f_\alpha^0$ is given by
\begin{align}
    G^* & \equiv \left . \frac{dG^{ss}}{df_\alpha} \right |_{f_\alpha^0} = \frac{4 \, \mathbf{K}^{ss}(f_\alpha^0) \rho_{eq}^0}{w \rho_0} \left . \frac{d\mathbf{M}}{df_\alpha} \right |_{f_\alpha^0} \nonumber \\
    &= \frac{12 \rho_{eq}^0 \ell^3 \mathbf{K}^{ss}(f_\alpha^0)}{\rho_0 w^4 f_\alpha^0 (1-f_\alpha^0)}.
    \label{eq:Gstar}
\end{align}
The sign of the slope of $G^{ss}(f_\alpha)$, and thus $f_\alpha^{ss}(G)$, is determined by the sign of $\mathbf{K}^{ss}(f_\alpha^0)$.

One can see that $f_\alpha$ is always stable to a small perturbation from steady state $\Delta f_\alpha \equiv f_\alpha - f_\alpha^{ss}(G)$ by writing Eq.~(\ref{eq:dfdta}) as
\begin{equation}
    \frac{df_\alpha}{dt} = \frac{\mathbf{K}^{dyn}(f_\alpha)}{\mathbf{K}^{ss}(f_\alpha)} \big [ G - G^{ss}(f_\alpha) \big ]. \label{eq:dfdta2a}
\end{equation}
For example, when $\mathbf{K}^{ss}$ is positive, and $\Delta f_\alpha$ is positive, then $G - G^{ss}(f_\alpha)$ will be negative, and the perturbation will decay.
For $f_\alpha$ near $f_\alpha^0$, the relaxation time $t^*$ of the perturbation can be obtained using $G - G^{ss}(f_\alpha) \approx - G^* \Delta f_\alpha$, giving
\begin{equation}
    \frac{1}{t^*} \equiv \frac{-1}{\Delta f_\alpha} \frac{df_\alpha}{dt} \approx \frac{\mathbf{K}^{dyn} \, G^*}{\mathbf{K}^{ss}} = \frac{12 \rho_{eq}^0 \ell^3 \mathbf{K}^{dyn}}{\rho_0 w^4 f_\alpha^0 (1-f_\alpha^0)}.
\end{equation}

To check the self-consistency of the assumption that the $\rhot_{eq}^j$ do not differ much from $\rho_{eq}^0$, used to obtain the simplified analytical solution, we require that the second term in Eq.~(\ref{eq:kt3}) is negligible with respect to $\rho_{eq}^0$, or
\begin{equation}
    \left | \frac{v_j \rho_0 \kappa_0^j}{\kapsq^j} \right | << \rho_{eq}^0 \label{eq:vjlim}
\end{equation}
for both steps $j = A$ and $B$.
We can write the expressions for the step velocities Eqs.~(\ref{eq:vas},\ref{eq:vbs}) as
\begin{align}
    v_A &= \frac{w G}{2} + \frac{2 \, \mathbf{M} \, \mathbf{K}^{dyn} \rho_{eq}^0}{\rho_0} \left [ \frac{G - G^{ss}}{G^{ss}} \right ],  \label{eq:vas2}\\
    v_B &= \frac{w G}{2} - \frac{2 \, \mathbf{M} \, \mathbf{K}^{dyn} \rho_{eq}^0}{\rho_0} \left [ \frac{G - G^{ss}}{G^{ss}} \right ]. \label{eq:vbs2}
\end{align}
The first term gives the steady-state velocity, and the second term gives the difference in velocity when $f_\alpha$ differs from $f_\alpha^{ss}$.
For the steady-state term, relation (\ref{eq:vjlim}) gives maximum growth rate magnitudes of
\begin{equation}
    |G| << \frac{2 \rho_{eq}^0 \kapsq^j}{w \rho_0 \kappa_0^j}
\end{equation}
for both steps $j = A$ and $B$.
For the dynamic term, relation (\ref{eq:vjlim}) gives maximum growth rate difference magnitudes of
\begin{equation}
    |G - G^{ss}| << \left | \frac{w \rho_0 [1 + R_0 (R_\alpha + R_\beta)] G^{ss} \kapsq^j}{2 D \mathbf{M} \rho_{eq}^0 (R_\alpha + R_\beta) \kappa_0^j} \right | .
\end{equation}
For the parameter ranges we consider, these limits on growth rate are many orders of magnitude larger than the growth rates relevant to this study, confirming the validity of the simplified analytical solution.
We have also checked that the exact solution obtained using the matrix equations Eqs.~(\ref{eq:MC}-\ref{eq:MB}) agrees with the simplified analytical solution.

\begin{figure}
\includegraphics[width=0.85\linewidth]{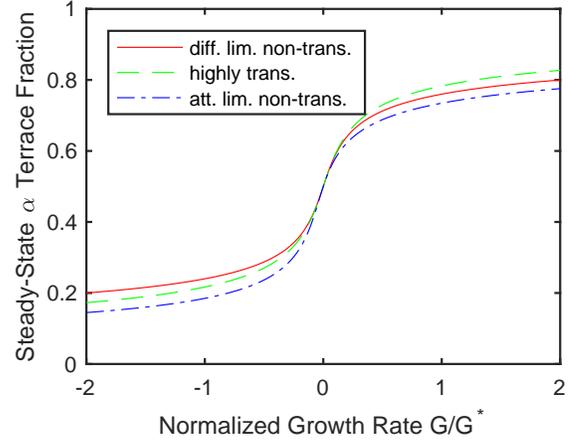}
\caption{Master curves of $f_\alpha^{ss}$ vs. $G/G^*$ for 3 cases: diffusion-limited kinetics with non-transparent steps, attachment-limited kinetics with non-transparent steps, and either kinetics with highly transparent steps. Parameter values used are given in Table \ref{tab:tableX}. \label{fig:master}}
\end{figure}

Figure~\ref{fig:master} shows the some examples of $f_\alpha^{ss}$ vs. $G/G^*$, calculated using the simplified analytical solution Eqs.~(\ref{eq:Dk}-\ref{eq:Gstar}) with parameter values given in Table \ref{tab:tableX}.
These correspond to some of the limiting cases discussed below.

\begin{table} 
\caption{ \label{tab:tableX} Parameter values used in BCF theory calculations for four sub-cases shown in Fig.~\ref{fig:master}. All used $w = 5.73 \times 10^{-8}$ m, $\rho_0 = 1.13 \times 10^{19}$ m$^{-2}$, $\ell = 9 \times 10^{-10}$ m, $\rho_{eq}^0 = 3.4 \times 10^{11}$ m$^{-2}$, $f_\alpha^0$ = 0.5.}
\begin{ruledtabular}
\begin{tabular}{ l || l | l | l | l }
Kinetics limited by: & diff. & diff. & attach. & attach. \\
\hline
Step transparency: & zero & high & zero & high \\
\hline
$D$ ~~~ (m$^2$/s) & $10^{-14}$ & $10^{-14}$ & $10^{-4}$ & $10^{-4}$  \\
$\kappa_+^A$ ~~ (m/s) & $10^2$ & $10^2$ & $10^2$ & $10^2$  \\
$\kappa_-^A$ ~~ (m/s) & $10^1$ & $10^1$ & $10^1$ & $10^1$  \\
$\kappa_0^A$ ~~ (m/s) & $0$ & $10^3$ & $0$ & $10^3$  \\
$\kappa_+^B$ ~~ (m/s) & $10^1$ & $10^1$ & $10^1$ & $10^1$  \\
$\kappa_-^B$ ~~ (m/s) & $10^0$ & $10^0$ & $10^0$ & $10^0$  \\
$\kappa_0^B$ ~~ (m/s) & $0$ & $10^3$ & $0$ & $10^3$  \\
\hline
$G^*$ ~~ (10$^{-3}$ ML/s) & $0.4$ & $1.2$ & $1.2$ & $1.2$ \\
\end{tabular}
\end{ruledtabular}
\end{table}

We next use the simplified analytical solution to consider two cases, in which the adatom kinetics on the terraces are limited by diffusion or by attachment/detachment at steps \cite{2020_Guin_PRL124_036101}.
For each, we consider the sub-cases of non-transparent or highly transparent steps, and examine the factors that determine the sign of $\mathbf{K}^{ss}$, and thus whether $f_\alpha^{ss}(G)$ has a positive or negative slope.
We finally consider a third case in which $\alpha$ and $\beta$ terraces have different limiting kinetics.

\subsection{Diffusion-limited kinetics}

In the diffusion-limited case, the first two terms are negligible in Eq.~(\ref{eq:Ra}) for $R_\alpha$ and in Eq.~(\ref{eq:Rb}) for $R_\beta$.
These expressions reduce to $R_\alpha = f_\alpha^{-1}$ and $R_\beta = (1 - f_\alpha)^{-1}$.
The coefficients $S_\alpha$ and $S_\beta$ become independent of $f_\alpha$.
The expression for $\mathbf{K}^{ss}$ is given by
\begin{equation}
    \mathbf{K}^{ss}(f_\alpha) \approx \big [ W_0^{dl} + W_1^{dl} f_\alpha (1 - f_\alpha) \big ]^{-1},
    \label{eq:Dk2}
\end{equation}
where we have introduced coefficients
\begin{align}
    W_0^{dl} &\equiv \frac{\kappa_0^B}{\kapsq^B} - \frac{\kappa_0^A}{\kapsq^A}, \\
    W_1^{dl} &\equiv \frac{\kappa_+^B}{\kapsq^B} + \frac{\kappa_-^B}{\kapsq^B} - \frac{\kappa_+^A}{\kapsq^A} - \frac{\kappa_-^A}{\kapsq^A} .
\end{align}
The expression for $\mathbf{K}^{dyn}$ becomes
\begin{equation}
    \mathbf{K}^{dyn} \approx \frac{D}{w [f_\alpha (1 - f_\alpha) + R_0]}. \label{eq:dfdta2}
\end{equation}

For the sub-case of non-transparent steps, with $\kappa_0^A = \kappa_0^B = 0$, we have $\kapsq^j = \kappa_+^j \kappa_-^j$ for both steps $j = A$ and $B$.
The expression for $\mathbf{K}^{ss}$ becomes
\begin{equation}
    \mathbf{K}^{ss}(f_\alpha) \approx
    \left [ f_\alpha (1 - f_\alpha) \left ( \frac{1}{\kappa_-^B} + \frac{1}{\kappa_+^B} - \frac{1}{\kappa_-^A} - \frac{1}{\kappa_+^A} \right ) \right ]^{-1}.
    \label{eq:Dk2nt}
\end{equation}
Here the smallest of the individual $\kappa_+^j$ or $\kappa_-^j$ tends to dominate and determine the sign of $\mathbf{K}^{ss}$.
The sign of $\mathbf{K}^{ss}$ is positive if the smallest coefficient is for the $B$ step, e.g. if the $B$ step has the higher ES barrier, so that $\kappa_-^B$ is smallest.
If there are no ES barriers, i.e. $\kappa_-^j = \kappa_+^j$, then the step with the smaller $\kappa_+^j$ determines the sign.
In this sub-case we have $R_0 = 0$, which simplifies Eq.~(\ref{eq:dfdta2}) for $df_\alpha/dt$.

For the sub-case of highly transparent steps, with $\kappa_0^j >> \kappa_+^j$ and $\kappa_-^j$, we have $\kapsq^j = \kappa_0^j (\kappa_+^j + \kappa_-^j)$ for both steps $j = A$ and $B$.
The expression for $\mathbf{K}^{ss}$ becomes a constant, independent of $f_\alpha$,
\begin{equation}
    \mathbf{K}^{ss} \approx
    \left ( \frac{1}{\kappa_-^B + \kappa_+^B} - \frac{1}{\kappa_-^A + \kappa_+^A} \right )^{-1}.
    \label{eq:Dk2t}
\end{equation}
Here the behavior just depends on the sums $\kappa_-^j + \kappa_+^j$ for each step.
It does not matter whether there are ES barriers;
the sign of $\mathbf{K}^{ss}$ is positive if $(\kappa_-^A + \kappa_+^A) > (\kappa_-^B + \kappa_+^B)$.

\subsection{Attachment-limited kinetics}

In the attachment-limited case, the final term is negligible in Eq.~(\ref{eq:Ra}) for $R_\alpha$ and in Eq.~(\ref{eq:Rb}) for $R_\beta$.
The coefficients $R_\alpha$ and $R_\beta$ become independent of $f_\alpha$.
The expression for $\mathbf{K}^{ss}$ is given by
\begin{equation}
    \mathbf{K}^{ss}(f_\alpha) \approx \big [ W_0^{al} + W_1^{al} (1 - 2f_\alpha) \big ]^{-1},
    \label{eq:Dk3}
\end{equation}
with coefficients
\begin{align}
    W_0^{al} & \equiv \frac{\kapsq^A - \kapsq^B + (\kappa_+^A + \kappa_-^A) \kappa_0^B - (\kappa_+^B + \kappa_-^B) \kappa_0^A}{(\kappa_+^B + \kappa_-^B) \kapsq^A + (\kappa_+^A + \kappa_-^A) \kapsq^B}, \label{eq:W0a} \\
    W_1^{al} & \equiv \frac{\kappa_+^B \kappa_+^A - \kappa_-^B \kappa_-^A}{(\kappa_+^B + \kappa_-^B) \kapsq^A + (\kappa_+^A + \kappa_-^A) \kapsq^B}. \label{eq:W1a}
\end{align}
The expression for $\mathbf{K}^{dyn}$ is independent of $f_\alpha$,
\begin{align}
    & \mathbf{K}^{dyn} \approx \left ( \left [ 
    \left ( \frac{\kappa_+^B}{\kapsq^B} + \frac{\kappa_-^A}{\kapsq^A} \right )^{-1} \right . \right . \nonumber \\
    &+ \left . \left . \left ( \frac{\kappa_-^B}{\kapsq^B} + \frac{\kappa_+^A}{\kapsq^A} \right )^{-1} \right ]^{-1}  
    +  \frac{\kappa_0^A}{\kapsq^A} + \frac{\kappa_0^B}{\kapsq^B} \right )^{-1} . \label{eq:dfdta3}
\end{align}
The diffusion coefficient $D$ does not enter into the solution for the attachment-limited case;
its role in the dynamics is taken by the combination of all the $\kappa$ coefficients given in Eq.~(\ref{eq:dfdta3}).
Since the denominators in Eqs.~(\ref{eq:W0a}-\ref{eq:W1a}) are always positive, the sign of $\mathbf{K}^{ss}$ is determined by the numerators.

For the sub-case of non-transparent steps, with $\kappa_0^A = \kappa_0^B = 0$, $\kapsq^j = \kappa_+^j \kappa_-^j$, the expressions for the coefficients in $\mathbf{K}^{ss}$ become
\begin{align}
    W_0^{al} & \equiv \frac{\kappa_+^A \kappa_-^A - \kappa_+^B \kappa_-^B}{(\kappa_+^B + \kappa_-^B) \kappa_+^A \kappa_-^A + (\kappa_+^A + \kappa_-^A) \kappa_+^B \kappa_-^B}, \label{eq:W0ant} \\
    W_1^{al} & \equiv \frac{\kappa_+^B \kappa_+^A - \kappa_-^B \kappa_-^A}{(\kappa_+^B + \kappa_-^B) \kappa_+^A \kappa_-^A + (\kappa_+^A + \kappa_-^A) \kappa_+^B \kappa_-^B}. \label{eq:W1ant}
\end{align}
This is the most complex sub-case.
Near $f_\alpha = 0.5$, the sign of $\mathbf{K}^{ss}$ is positive if $\kappa_+^B \kappa_-^B <  \kappa_+^A \kappa_-^A$.
At $f_\alpha > 0.5$, if the steps have normal ES barriers with $\kappa_-^j < \kappa_+^j$, the $W_1^{al}$ term will favor a negative sign.
Thus the sign of $\mathbf{K}^{ss}$ can change with $f_\alpha$.
The expression for $\mathbf{K}^{dyn}$ becomes
\begin{equation}
    \mathbf{K}^{dyn} \approx 
    \left ( \frac{1}{\kappa_-^B} + \frac{1}{\kappa_+^A} \right )^{-1} +  \left ( \frac{1}{\kappa_+^B} + \frac{1}{\kappa_-^A} \right )^{-1}. \label{eq:dfdta3nt}
\end{equation}
The dynamic coefficient has an interesting form, dominated by the terrace with the largest value of the \textit{smallest} attachment coefficient at its edges.

For the sub-case of highly transparent steps, with $\kappa_0^j >> \kappa_+^j$ and $\kappa_-^j$, $\kapsq^j = \kappa_0^j (\kappa_+^j + \kappa_-^j)$, the expression for $\mathbf{K}^{ss}$ becomes a constant identical to that for diffusion-limited kinetics with highly transparent steps,
\begin{equation}
    \mathbf{K}^{ss} \approx
    \left ( \frac{1}{\kappa_-^B + \kappa_+^B} - \frac{1}{\kappa_-^A + \kappa_+^A} \right )^{-1}.
    \label{eq:Dk3t}
\end{equation}
As before, the steady-state behavior just depends on the sums $\kappa_-^j + \kappa_+^j$ for each step.
The dynamics still differs from the diffusion-limited case, since the expression for $\mathbf{K}^{dyn}$ differs from Eq.~(\ref{eq:dfdta2}),
\begin{equation}
    \mathbf{K}^{dyn} \approx \left ( 
    \frac{1}{\kappa_-^B + \kappa_+^B} + \frac{1}{\kappa_-^A + \kappa_+^A} \right )^{-1} . \label{eq:dfdta3t}
\end{equation}

\subsection{Mixed kinetics}

The limits considered above assume that both terraces have the same kinetics, either diffusion- or attachment-limited, and that both steps have the same transparency, either zero or high. 
Because the attachment coefficients can be different for each step type, other limiting cases are possible.
Here we consider the limit in which the $\kappa_+^A$ coefficient is much larger than the other five $\kappa_x^j$, so that the $A$ step has a high ES barrier, with $\kappa_-^A + \kappa_0^A << D / w f_\alpha$ (the $A$ step is non-transparent).
We also assume that $\kappa_-^B << \kappa_+^B \kappa_0^B / (\kappa_+^B + \kappa_0^B)$ so that the $B$ step also has a high ES barrier.
In this case we have $\kapsq^A = \kappa_+^A (\kappa_-^A + \kappa_0^A)$ and $\kapsq^B = \kappa_+^B \kappa_0^B$.
The second and third terms in Eq.~(\ref{eq:Ra}) are negligible, giving $R_\alpha = (w/D)(\kappa_-^A + \kappa_0^A)$.
The second term in Eq.~(\ref{eq:Rb}) is negligible, giving $R_\beta = [ D/(w \kappa_0^B) + (1 - f_\alpha)]^{-1}$.
The second terms in Eqs.~(\ref{eq:Sa}) and (\ref{eq:Sb}) are negligible, giving $S_\alpha = w^2 f_\alpha / (2D)$, $S_\beta = (w/2)(1-f_\alpha)/[D/w + (1-f_\alpha)\kappa_0^B]$.
The first terms in Eqs.~(\ref{eq:R0}) and (\ref{eq:S0}) are negligible, giving $R_0 = D/(w \kappa_+^B)$, $S_0 = -w/(2 \kappa_+^B)$.
This results in expressions
\begin{equation}
    \mathbf{K}^{ss}(f_\alpha) \approx \left [ 
    \frac{1}{\kappa_+^B} + \frac{(1 - 2 f_\alpha)}{\kappa_0^B} - \frac{w f_\alpha (1-f_\alpha)}{D} \right ]^{-1},
    \label{eq:Dkm}
\end{equation}
\begin{align}
    & \frac{df_\alpha}{dt} \approx \left [ 
    \frac{D}{\kappa_+^B} + \frac{D}{\kappa_0^B} + w(1-f_\alpha) \right ]^{-1} \times \nonumber \\ 
    & \left ( G \left [ 
    \frac{D}{\kappa_+^B} + \frac{D (1 - 2 f_\alpha)}{\kappa_0^B} - w f_\alpha (1-f_\alpha) \right ] - \right . \nonumber \\
    & \left . \frac{4 D \rho_{eq}^0}{w \rho_0} \left ( \frac{\ell}{w} \right )^3 \left [ \left ( \frac{1 - f_\alpha^0}{1 - f_\alpha} \right )^3 - \left ( \frac{f_\alpha^0}{f_\alpha} \right)^3 \right ] \right ), \label{eq:dfdtm}
\end{align}
\begin{align}
    & G^{ss} \approx \frac{4 D \rho_{eq}^0}{w \rho_0} \left ( \frac{\ell}{w} \right )^3 \left [ \left ( \frac{1 - f_\alpha^0}{1 - f_\alpha} \right )^3 - \left ( \frac{f_\alpha^0}{f_\alpha} \right)^3 \right ] \times \nonumber \\
    & \left [ 
    \frac{D}{\kappa_+^B} + \frac{D (1 - 2 f_\alpha)}{\kappa_0^B} - w f_\alpha (1-f_\alpha) \right ]^{-1}. \label{eq:Gssm}
\end{align}
Even though $\kappa_+^A$ has the largest value, the sign of $\mathbf{K}^{ss}$ can be negative depending upon the relative size of the terms in Eq.~(\ref{eq:Dkm}).
It will be negative near $f_\alpha = 0.5$ for $D/\kappa_+^B < w/4$.
If $\kappa_0^B$ is small, it can become negative for $f_\alpha > 0.5$.

\section{Comparison of BCF theory to X-ray measurements}

The BCF model predicts the dependence of the steady-state terrace fraction on growth rate $f_\alpha^{ss}(G)$, as well as the dynamics of the transitions when $G$ is changed.
We can compare calculated values to our measurements to understand the implications for the physics in the model, such as the differences between adatom attachment kinetics at $A$ and $B$ steps.

In the general model, e.g. Eqs.~(\ref{eq:MCt})-(\ref{eq:ptqt}), there are 14 fundamental variables ($F$, $\tau$, $\rho_0$, $w$, $D$, $\rho_{eq}^0$, $\ell$, $f_\alpha^0$, and the six $\kappa_x^j$).
In the simplified analytical solution presented in Section IV.D., four variables enter only through two combinations  ($G = F - \rho_{eq}^0/\tau$, and $\rho_{eq}^0 \ell^3$), leaving 12 independent variables.
We control or directly determine $G$, $\rho_0$, and $w$, leaving 9 unknown quantities ($D$, $\rho_{eq}^0 \ell^3$, $f_\alpha^0$, and the six combinations of the $\kappa_x^j$) to be determined or constrained by the measurements.
This is a challenge because we have only 6 measured quantities (four steady-state $\alpha$ terrace fractions $f_\alpha^{ss}$ at different growth rates $G$, and two relaxation times for transitions in $G$.)

As we have seen, in some limits the number of effective parameters is smaller, since only certain combinations of $D$ and the $\kappa_x^j$ enter the solutions.
The diffusion-limited kinetics solutions reduce these 7 to 4 combinations, leaving a total of 6 unknown quantities.
The sub-cases of non-transparent or highly transparent steps reduce the number of effective parameters by one or two more.
The attachment-limited kinetics solutions reduce these 7 to 2 combinations, leaving a total of 4 unknown quantities.
The highly transparent sub-case reduces this by one.
The mixed kinetics solution has a total of 4 unknown quantities, $D/\kappa_+^B$, $D/\kappa_0^B$, $D \rho_{eq}^0 \ell^3$, and $f_\alpha^0$. 

To calculate BCF model results to compare with the experimental conditions, we assume that the only parameter affected by the TEGa supply rate is the deposition flux $F$, and that the only parameter affected by the carrier gas composition (0\% or 50\% H$_2$) is the adatom lifetime $\tau$, and that these enter only through the net growth rates $G$ given in Table \ref{tab:tableZ} for each condition, as determined in Appendix C.
We use the known values $\rho_0 = 2 a^{-2} / \sqrt{3} = 1.13 \times 10^{19}$ m$^{-2}$ and $w = c / \sin(0.52^\circ) = 5.73 \times 10^{-8}$ m, where $a = 3.20 \times 10^{-10}$ m and $c = 5.20 \times 10^{-10}$ m are the lattice parameters of GaN at the growth temperature \cite{2000_Reeber_JMaterRes15_40}.

\begin{table} 
\caption{ \label{tab:tableZ} Comparison of measured values to best fit calculated from the simplified analytical solution of the BCF model, using parameters given in Table \ref{tab:tableY}.}
\begin{ruledtabular}
\begin{tabular}{ l | l || l | l | l }
Cond./ & $G$ & & Measured & Best\\
Trans. & (ML/s) & & Value & Fit \\
\hline
$1$ & -0.0018 & $f_\alpha^{ss}$ & $0.111 \pm 0.013$ & $0.136$\\
$2$ & 0.0000 & $f_\alpha^{ss}$ & $0.461 \pm 0.018$ & $0.440$\\
$3$ & 0.0109 & $f_\alpha^{ss}$ & $0.811 \pm 0.014$ & $0.836$\\
$4$ & 0.0127 & $f_\alpha^{ss}$ & $0.867 \pm 0.011$ & $0.847$\\
\hline
$1$ to $2$ &  & $t_{rel}$ & $2177 \pm 218$ s & $2478$ \\
$2$ to $4$ &  & $t_{rel}$ & $341 \pm 34$ s & $331$ \\
\hline
 & & $\chi^2$ & $-$ & 13.6\\
\end{tabular}
\end{ruledtabular}
\end{table}

\begin{table} 
\caption{ \label{tab:tableY} Parameter values used in simplified analytical BCF model calculations given in Table~\ref{tab:tableZ}.}
\begin{ruledtabular}
\begin{tabular}{ c || c | c | c }
Parameter & Best-fit & Best-fit & Units\\
& Solution \#1 & Solution \#2 & \\
\hline
$D$ & $1.00 \times 10^{-4}$ & $1.00 \times 10^{-9}$ & (m$^2$/s)  \\
$\kappa_+^A$ & (large) & (large) & (m/s)  \\
$\kappa_+^B$ & $5.27 \times 10^3$ & $5.27 \times 10^{-2}$ & (m/s)  \\
$\kappa_-^A$ & $\approx 0$ & $\approx 0$ & (m/s)  \\
$\kappa_-^B$ & $\approx 0$ & $\approx 0$ & (m/s)  \\
$\kappa_0^A$ & $\approx 0$ & $\approx 0$ & (m/s)  \\
$\kappa_0^B$ & $9.45 \times 10^3$ & $9.45 \times 10^{-2}$ & (m/s)  \\
$\rho_{eq}^0 \ell^3$ & $3.31 \times 10^{-19}$ & $3.31 \times 10^{-14}$ & (m)  \\
$f_\alpha^0$ & $0.440$ & $0.440$ & $-$ \\
\end{tabular}
\end{ruledtabular}
\end{table}

We searched the space of the 9 unknown quantities of the simplified analytical solution to find the best fit to the measured quantities.
Table \ref{tab:tableZ} compares the six measured quantities (four steady-state values of $f_\alpha^{ss}$ and two relaxation times $t_{rel}$ following growth rate transitions) to the best-fit values calculated from the BCF model.
The best fit was determined by minimizing the goodness-of-fit parameter $\chi^2 \equiv \sum [(y_i - y_i^{calc})/\sigma_i]^2$, where the $y_i$ and $\sigma_i$ are the six measured quantities and their uncertainties.
To estimate the uncertainties in the $f_\alpha^{ss}$, we multiplied those obtained in the fits to the 3H(T1) reconstruction by a factor of 4, to account for the uncertainties in the atomic coordinates used.
We estimated the uncertainty in the $t_{rel}$ to be 10\%.
We found a family of equivalent solutions giving essentially the same results and the same minimum $\chi^2$.
Two examples with different parameter value sets, denoted \#1 and \#2, are shown in Table~\ref{tab:tableY}.
For this region of parameter space, the values of several of the parameters could be varied with no significant effect, as long as they were sufficiently large or close to zero, as indicated in Table~\ref{tab:tableY}.
These best-fit solutions to the simplified analytical model correspond to the mixed kinetics limit described above.

\begin{table} 
\caption{ \label{tab:tableL} Comparison of measured values to those calculated from limiting cases of the BCF model, using parameters given in Table \ref{tab:tableLP}.}
\begin{ruledtabular}
\begin{tabular}{ l | l || l | l | l | l | l }
Cond./ & $G$ & & Measured & Diff. & Attach. & Mixed\\
Trans. & (ML/s) & & Value & Ltd. & Ltd. & Kin. \\
\hline
$1$ & -0.0018 & $f_\alpha^{ss}$ & $0.111 \pm 0.013$ & $0.159$ & $0.154$ & $0.136$  \\
$2$ & 0.0000 & $f_\alpha^{ss}$ & $0.461 \pm 0.018$ & $0.389$ & $0.410$& $0.440$  \\
$3$ & 0.0109 & $f_\alpha^{ss}$ & $0.811 \pm 0.014$ & $0.863$ & $0.830$& $0.836$  \\
$4$ & 0.0127 & $f_\alpha^{ss}$ & $0.867 \pm 0.011$ & $0.871$ & $0.838$& $0.847$  \\
\hline
$1$ to $2$ &  & $t_{rel}$ & $2177 \pm 218$ s & $2204$ & $2958$ & $2478$  \\
$2$ to $4$ &  & $t_{rel}$ & $341 \pm 34$ s & $337$ & $251$ & $331$  \\
\hline
 & & $\chi^2$ & $-$ & 42.5  & 46.7 & 13.6 \\
\end{tabular}
\end{ruledtabular}
\end{table}

\begin{table} 
\caption{ \label{tab:tableLP} Best-fit parameter values for the three limiting cases of the BCF model, for fits shown in Table~\ref{tab:tableL} and Fig.~\ref{fig:fss_exp}.}
\begin{ruledtabular}
\begin{tabular}{ c }
Diffusion-limited kinetics \\
\end{tabular}
\begin{tabular}{ c || c | c }
Parameter & Value & Units\\
\hline
$D W_0^{dl}$ & $2.29 \times 10^{-9}$ & (m)  \\
$D W_1^{dl}$ & $-5.27 \times 10^{-9}$ & (m)  \\
$D \rho_{eq}^0 \ell^3$ & $6.18 \times 10^{-24}$ & (m$^3$/s)  \\
$f_\alpha^0$ & $0.389$ & $-$ \\
\end{tabular}
\begin{tabular}{ c }
Attachment-limited kinetics \\
\end{tabular}
\begin{tabular}{ c || c | c }
Parameter & Value & Units\\
\hline
$\mathbf{K}^{dyn} W_0^{al}$ & $9.46 \times 10^{-2}$ & $-$  \\
$\mathbf{K}^{dyn} W_1^{al}$ & $6.30 \times 10^{-2}$ & $-$ \\
$\mathbf{K}^{dyn} \rho_{eq}^0 \ell^3$ & $4.17 \times 10^{-16}$ & (m$^2$/s)  \\
$f_\alpha^0$ & $0.410$ & $-$  \\
\end{tabular}
\begin{tabular}{ c }
Mixed kinetics \\
\end{tabular}
\begin{tabular}{ c || c | c }
Parameter & Value & Units\\
\hline
$D / \kappa_+^B$ & $1.90 \times 10^{-8}$ & (m)  \\
$D / \kappa_0^B$ & $1.06 \times 10^{-8}$ & (m)  \\
$D \rho_{eq}^0 \ell^3$ & $3.31 \times 10^{-23}$ & (m$^3$/s)  \\
$f_\alpha^0$ & $0.440$ & $-$ \\
\end{tabular}
\end{ruledtabular}
\end{table}

To understand how well the measurements constrain the model parameters and the physics underlying them, we have also searched for the best fit for each of the three limiting cases.
For the diffusion-limited case, the best fit occurs with the parameter $R_0$ negligible in Eq.~(\ref{eq:dfdta2}), so that only four combinations of unknown quantities are needed to specify the solution, as in the attachment-limited and mixed kinetics cases.
Table~\ref{tab:tableL} compares the results of these fits, and Table~\ref{tab:tableLP} summarizes the best-fit values of the four quantities obtained for each limiting case.
We have also plotted the curves of $f_\alpha^{ss}(G)$ corresponding to these fits with the experimental points in Fig.~\ref{fig:fss_exp}.
It is clear that the mixed kinetics limit gives a significantly better fit.

To interpret the combined parameters obtained from the fits, it is useful to estimate the adatom diffusivity $D$ and equilibrium adatom density $\rho_{eq}^0$.
\textit{Ab initio} calculations of the activation energy for Ga diffusion on the Ga-terminated (0001) surface have given values of $\Delta H_m = 0.4$~eV \cite{1998_Zywietz_APL73_487} and $\Delta H_m = 0.5$~eV \cite{2017_Chugh_PCCP19_2111}, and similar values have been obtained for 3d transition metal adatoms \cite{2011_Gonzalez-Hernandez_JAP110_083712}.
An estimate based on spatial correlations in the surface morphology of GaN films grown at two temperatures gave $\Delta H_m = 1.6 \pm 0.5$~eV \cite{2014_Koleske_JCrystGrowth391_85}.
If we estimate the diffusivity from the \textit{ab initio} calculations using $D = a^2 \nu \exp(\Delta S_m / k) \exp(-\Delta H_m / kT)$ \cite{1989_Shewmon_2ndEd_DiffusioninSolids}, with $a = 3.2 \times 10^{-10}$~m, $\nu = 10^{14}$~s$^{-1}$, $\Delta S_m = 0$, and $\Delta H_m = 0.4$~eV, we obtain $D = 1.4 \times 10^{-7}$ m$^2$/s at $T = 1073$~K.
In addition, the surface morphology analysis \cite{2014_Koleske_JCrystGrowth391_85} indicated a cross-over at $T = 1073$~K from surface diffusion transport to evaporation/condensation transport at a length scale of $\lambda = 1.5 \times 10^{-6}$~m for OMVPE growth with H$_2$ present in the carrier gas.
Thus the adatom lifetime $\tau$ can be estimated as $\tau = \lambda^2/D = 1.7 \times 10^{-5}$~s under these conditions.
Using our observed negative net growth rate for $F = 0$ of $G = -\rho_{eq}^0/(\rho_0 \tau) = -0.00184$~ML/s, this gives a value for the equilibrium adatom density of $\rho_{eq}^0 = 3.4 \times 10^{11}$~m$^{-2}$.
Using these estimates for $D$ and $\rho_{eq}^0$, the parameters obtained from the mixed kinetics fit imply kinetic coefficients of $\kappa_+^B = 7.4$~m/s and $\kappa_0^B = 13$~m/s, and a step repulsion length of $\ell = 9 \times 10^{-10}$~m. 
The example calculations shown in Figs.~\ref{fig:dfdt1}-\ref{fig:fvst1} correspond to these parameter values.

\section{Discussion and Conclusions}

Although it has not been possible using scanning-probe microscopy to observe the orientation difference of $\alpha$ and $\beta$ terraces on vicinal basal plane surfaces of HCP-type systems, our results show that this difference is robustly revealed by surface X-ray scattering.
\textit{In situ} X-ray measurements during growth can determine the fraction covered by each terrace, and thus distinguish the dynamics of $A$ and $B$ steps.
While the CTR calculations presented here are for wurtzite-structure GaN, this method applies to many other HCP-type systems with a $6_3$ screw axis, including other compound semiconductors, as well as one third of the crystalline elements and many more complex crystals.

The BCF model we have developed makes detailed predictions for the behavior of the $\alpha$ terrace fraction $f_\alpha$ at steady-state and during transients, in terms of surface properties such as the adatom diffusivity $D$ and step kinetic coefficients $\kappa_x^j$.
In particular, the steady-state fraction $f_\alpha^{ss}$ is predicted to depend only on the net growth rate $G = (F - \rho_{eq}^0 \tau)/\rho_0$, rather than individually on the deposition rate $F$ or the adatom lifetime $\tau$.
The positive or negative slope of $f_\alpha^{ss}(G)$ is determined by the sign of a combined kinetic parameter $\mathbf{K}^{ss}$.
For diffusion- or attachment-limited kinetics, whether non-transparent or highly transparent, the sign of $\mathbf{K}^{ss}$ is determined solely by the values of the attachment parameters $\kappa_+^j$ and $\kappa_-^j$ at the two types of steps $j = A$ or $B$, independent of the transmission coefficients $\kappa_0^j$ or $D$.
This is unlike the mixed-kinetics case, where the values of $\kappa_0^B$ and $D$ play a role in determining the sign.

Our primary experimental result, the positive slope of $f_\alpha^{ss}(G)$, determines the basic nature of the adatom attachment kinetics at $A$ and $B$ steps.
In general, this slope is positive if the $A$ step attachment coefficients $\kappa_+^A$ and/or $\kappa_-^A$ are larger than the $B$ step attachment coefficients $\kappa_+^B$ and/or $\kappa_-^B$.
While the same general shape of $f_\alpha^{ss}(G)$ can be obtained by many combinations of the parameters in the BCF model that have faster $A$ step kinetics, the best fit to our steady-state and dynamics measurements is obtained in a specific mixed kinetics limit.
Assuming that both terraces are have either diffusion-limited or attachment-limited kinetics gives significantly worse fits.
The agreement with the mixed kinetic limit indicates much faster attachment kinetics at the $A$ step than the $B$ step, with $\kappa_+^A >> \kappa_+^B$.
It indicates that both $A$ and $B$ steps have standard positive ES barriers, with adatom attachment from below significantly faster than from above, for the same supersaturation. 
This limit also indicates that the $A$ step is non-transparent.
The fit also gives a value for $f_\alpha^0$ differing slightly from the symmetrical value of 1/2.

In evaluating the values of the step kinetic coefficients, the $5^\circ$ rotation of the step azimuth away from $[0 1 \overline{1} 0]$ is potentially important, since it determines the average kink spacing on the steps to be $b/2 \tan 5^\circ = 3.2$~nm.
We expect that this relatively small kink spacing will tend to produce higher values of the attachment coefficients $\kappa_+^j$ and $\kappa_-^j$ and lower values of the transmission coefficients $\kappa_0^j$, since attachment occurs when adatoms at a step diffuse along it to a kink before leaving the step \cite{2007_Ranguelov_PRB75_245419}.

Our result that $A$ steps have higher attachment coefficients than $B$ steps disagrees with most predictions in the literature \cite{2010_Zaluska-Kotur_JNoncrystSolids356_1935,2011_Zaluska-Kotur_JAP109_023515,2017_Chugh_ApplSurfSci422_1120,2017_Xu_JChemPhys146_144702,2013_Turski_JCrystGrowth367_115,2020_Akiyama_JCrystGrowth532_125410,2020_Akiyama_JJAP59_SGGK03}.
It agrees with the original proposal \cite{1999_Xie_PRL82_2749} based on a specific bond-counting argument and analogy with experiments on GaAs (111) surfaces. 
Such predictions depend on the environmental conditions assumed, and several of these studies focused on MBE conditions.
For example, arguments regarding dangling bonds at steps \cite{1999_Xie_PRL82_2749,2013_Turski_JCrystGrowth367_115} depend on how they are passivated by the environment, including the effects of very high or low V/III ratios \cite{2017_Pristovsek_physstatsolb254_1600711} and the presence of NH$_3$ or H$_2$.
Likewise, KMC studies \cite{2011_Zaluska-Kotur_JAP109_023515,2010_Zaluska-Kotur_JNoncrystSolids356_1935,2017_Xu_JChemPhys146_144702,2017_Chugh_ApplSurfSci422_1120} typically make assumptions about bonding that determine the rates of atomic-scale processes at steps.
Detailed \textit{ab initio} predictions of ES barriers and adsorption energies at steps under MBE conditions \cite{2020_Akiyama_JCrystGrowth532_125410,2020_Akiyama_JJAP59_SGGK03} show that they depend strongly on the amount of excess Ga on the surface.
In future theoretical work, it would be useful to consider the specific step-edge structures associated with the OMVPE environment with the 3H(T1) reconstruction found here.

We have demonstrated this X-ray method using micron-scale X-ray beams to illuminate regions of surface with a well-defined step azimuth, which is critical for success.
With current synchrotron X-ray sources, it is convenient to increase the signal rate using wide-energy-bandwidth pink beam.
The higher brightness synchrotron sources soon to come online worldwide will make it possible to perform these experiments with highly monochromatic beams, greatly increasing the in-plane resolution of the CTR measurements.

\begin{acknowledgments}
  Work supported by the U.S Department of Energy (DOE), Office of Science, Office of Basic Energy Sciences, Materials Science and Engineering Division.
  Experiments performed at the Advanced Photon Source beamline 12ID-D, a DOE Office of Science user facility.
\end{acknowledgments}

\appendix
\section{Chemical potentials in OMVPE}

To calculate the CTR intensities to fit to the experimental profiles, we need the coordinates $\mathbf{r}_{jkn}$ of the atoms in the reconstructed layers.
The relaxed coordinates and free energies of various surface reconstructions for GaN (0001) in the OMVPE environment containing NH$_3$ and H$_2$ have been calculated \cite{2002_VandeWalle_PRL88_066103, 2012_Walkosz_PRB_85_033308}, leading to a phase diagram that can be expressed in terms of the chemical potentials of Ga and NH$_3$ \cite{2002_VandeWalle_JVacSciTechnolB20_1640,2012_Walkosz_PRB_85_033308}.
In this section we estimate these chemical potentials from the conditions in our experiments, to locate the appropriate region of the phase diagram and identify the predicted reconstructions in this region.

Figure~\ref{fig:recon_phase} shows the predicted surface phase diagram \cite{2012_Walkosz_PRB_85_033308}. 
The vertical axis is the chemical potential of NH$_3$ relative to its value at $T = 0$~K.
This can be expressed as
\begin{align}
    \Delta \mu_{NH_3}(T) &\equiv \mu_{NH_3}(T) - \mu_{NH_3}(0) \nonumber \\
    &= G_{NH_3}^{\circ}(T) - G_{NH_3}^{\circ}(0) + kT \log p_{NH_3},
    \label{eq:DmuNH3}
\end{align}
where $G_{NH_3}^{\circ}$ is the free energy of NH$_3$ gas at a pressure of 1 bar obtained from thermochemical tables \cite{1998_Chase_JPCRDMono9_NIST-JANAF}, 
and $p_{NH_3}$ is the partial pressure of NH$_3$ in the experiment.
These can be evaluated at the experimental conditions.
For $T = 1073$~K, the tables give $G_{NH_3}^{\circ}(T) - G_{NH_3}^{\circ}(0) = -2.1$~eV.
Thus for $p_{NH_3} = 0.04$~bar, one obtains 
$\Delta \mu_{NH_3}(T) = -2.4$~eV.

The horizontal axis in Fig.~\ref{fig:recon_phase} is the chemical potential of Ga relative elemental liquid Ga.
This can be related to the activity of N$_2$ using
\begin{align}
    \Delta \mu_{Ga} &\equiv \mu_{Ga}(T) - \mu_{Ga}^{liq}(T) \nonumber \\ 
    &= \Delta G_{f}^{GaN}(T) - 0.5 kT \log a_{N_2},
    \label{eq:DmuGa}
\end{align}
where $\Delta G_{f}^{GaN}$ is the free energy of formation of GaN from liquid Ga and N$_2$ gas at 1 bar, and $a_{N_2}$ is the activity (effective partial pressure) of N$_2$.

\begin{figure}
\includegraphics[width=3.0in]{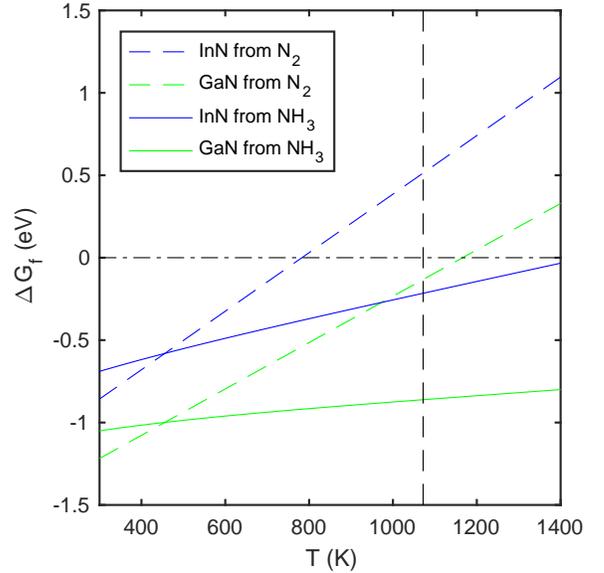}
\caption{\label{fig:DGf} Free energy of formation as a function of temperature of InN and GaN from the liquid metals and either vapor N$_2$ or NH$_3$ at 1 bar \cite{1998_Chase_JPCRDMono9_NIST-JANAF,1996_Ambacher_JVSTB14_3532}. In the case of NH$_3$, this includes formation of H$_2$ at 1 bar.}
\end{figure}

In OMVPE, a chemically active precursor such as ammonia is typically used to provide the high nitrogen activity required to grow group III nitrides.
The need for this can be seen in Fig.~\ref{fig:DGf}, which shows the free energies of the reactions to form GaN and InN from the condensed metallic elements and either vapor N$_2$ or vapor NH$_3$ at 1 bar \cite{1998_Chase_JPCRDMono9_NIST-JANAF,1996_Ambacher_JVSTB14_3532}.
At typical temperatures used for growth of high quality single crystal films at high rates (e.g. 1000~K for InN, 1300~K for GaN), the formation energy from N$_2$ is positive, indicating that the nitride is not stable and cannot be grown from N$_2$ at 1 bar.
In contrast, the formation energies of the nitrides (plus H$_2$ at 1 bar) from the metals and NH$_3$ are negative at all relevant growth temperatures,
indicating that growth from 1 bar of NH$_3$ is possible.

However, actual OMVPE conditions do not correspond with equilibrium, because the very high partial pressures of N$_2$ and/or H$_2$ that would correspond to equilibrium with NH$_3$ at these temperatures are not allowed to accumulate.
Thus, while formation of InN and GaN from NH$_3$ is energetically favored under OMVPE conditions, decomposition of these nitrides into N$_2$ is also energetically favored.
This metastability is manifested in the oscillatory growth and decomposition of InN that has been observed~\cite{2008_Jiang_PRL101_086102}.
Thus the kinetics of the reaction steps that determine the nitrogen activity at the growth surface are critical to understanding and controlling OMPVE growth of metastable nitrides.

In previous work we have measured the trimethylindium (TMI) partial pressures required to condense InN and elemental In onto GaN (0001) \cite{2008_Jiang_PRL101_086102}.
They can be analyzed to give experimentally determined values for the effective surface nitrogen activity arising from NH$_3$ under OMVPE conditions.
The experiments were carried out using a very similar growth chamber~\cite{1999_Stephenson_MRSBull24_21} as that used for the {\it in-situ} X-ray studies described below,
using the same a total pressure of 0.267 bar, and the same NH$_3$ and carrier flows (2.7 standard liters per minute (slpm) NH$_3$ and 1.1 slpm N$_2$ in the group V channel, 0.9 slpm N$_2$ carrier gas for TMI in the group III channel). 
We have performed chamber flow modeling to calculate the equivalent TMI and NH$_3$ partial pressures $p_{TMI}$ and $p_{NH_3}$ above the center of the substrate surface as a function of inlet flows.
At typical growth temperatures, an inlet flow of 0.184 $\mu$mol/min TMI corresponds to $p_{TMI} = 1.22 \times 10^{-6}$ bar, and 
an inlet flow of 2.7 slpm NH$_3$ corresponds to $p_{NH_3} = 0.040$ bar.

\begin{figure}
\includegraphics[width=3.0in]{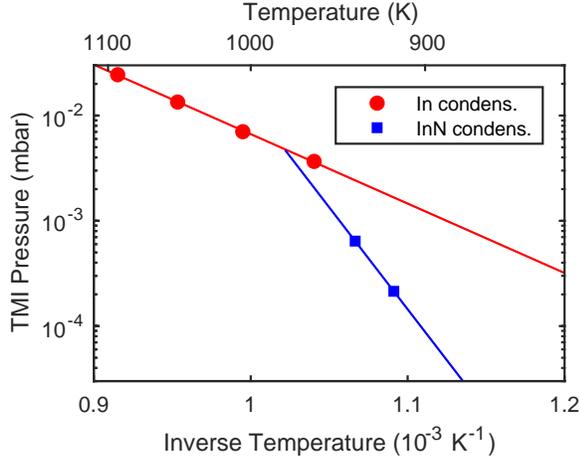}
\caption{\label{fig:InN} Observed phase boundaries for condensation onto GaN (0001) of relaxed epitaxial InN (blue squares) and liquid elemental In (red circles) at $p_{NH_3} = 0.040$~bar \cite{2008_Jiang_PRL101_086102}.}
\end{figure}

Figure~\ref{fig:InN} shows the $p_{TMI}$-$T$ boundaries determined by {\it in-situ} X-ray fluorescence and diffraction measurements for initial condensation of elemental In liquid or crystalline InN onto a GaN (0001) surface at $p_{NH_3} = 0.040$ bar \cite{2008_Jiang_PRL101_086102}. 
At TMI partial pressures above the boundaries shown, the condensed phases nucleate and grow on the surface;
at lower $p_{TMI}$, the condensed phases evaporate. 
The InN and In condensation boundaries intersect at 979 K. 

A relationship between the nitrogen and indium activities at the InN condensation boundary can be obtained from the equilibrium
\begin{equation}
{\rm In}_{vap} + \frac12 {\rm N}_2 \leftrightarrow {\rm InN}_{sol},
\end{equation}
which gives the chemical potential $\mu_i$ expression
\begin{equation}
\mu_{In} + \frac12 \mu_{N_2} = \mu_{InN},
\end{equation}
and the activity $a_i$ expression
\begin{equation}
kT \log a_{In} + \frac12 kT \log a_{N_2} = \Delta G_f^{InN}(T),
\label{eq:aIn}
\end{equation}
where $\Delta G_f^{InN}(T)$ is the formation energy of InN from liquid In and N$_2$ at 1 bar shown in Figure~\ref{fig:InN}. 
We assume that the activity of In relative to liquid In at the InN boundary is equal to the ratio $a_{In} = p_{TMI}^{InN} / p_{TMI}^{In}$, giving
\begin{equation}
kT \log a_{In} = kT \log p_{TMI}^{InN} - kT \log p_{TMI}^{In}
\end{equation}
at the experimental condition, $p_{NH_3} = 0.040$~bar. 
Equation~(\ref{eq:aIn}) can then be used to obtain the nitrogen activity relative to 1 bar (i.e. effective partial pressure of N$_2$ in bar) for $p_{NH_3} = 0.040$~bar. 

\begin{table} 
\caption{ \label{tab:aN2} Evaluation of N$_2$ activity and $\Delta \mu_{Ga}$ at the GaN surface under OMVPE conditions. Formation energies of GaN and InN are from elements at standard conditions. TMI pressures at InN and In condensation boundaries are for $p_{NH_3} = 0.04$~bar. Calculated $a_{N_2}$ and $\Delta \mu_{Ga}$ are thus also for $p_{NH_3} = 0.04$~bar.}
\begin{ruledtabular}
\begin{tabular}{ c || c }
Quantity & Value as $f(T)$~(K)\\
 & (eV) \\
\hline
$\Delta G_f^{GaN}$ \cite{1996_Ambacher_JVSTB14_3532} & $-1.64 + 1.41 \times 10^{-3} T$ \\
$\Delta G_f^{InN}$ \cite{1996_Ambacher_JVSTB14_3532} & $-1.39 + 1.78 \times 10^{-3} T$ \\
$kT \log p_{TMI}^{InN}$ \cite{2008_Jiang_PRL101_086102} & $-1.309 + 0.88 \times 10^{-3} T$ \\
$kT \log p_{TMI}^{In}$ \cite{2008_Jiang_PRL101_086102} & $-3.843 + 3.47 \times 10^{-3} T$ \\
\hline
$kT \log a_{In}$ & $-2.534 + 2.59 \times 10^{-3} T$ \\
$= kT \log p_{TMI}^{InN} - kT \log p_{TMI}^{In}$ & \\
\hline
$kT \log a_{N_2}$ & $2.288 - 1.63 \times 10^{-3} T$ \\
$= 2(\Delta G_f^{InN} - kT \log a_{In})$ & \\
\hline
$\Delta \mu_{Ga}$ & $-2.784 + 2.225 \times 10^{-3} T$ \\
$= \Delta G_f^{GaN} - 0.5 kT \log a_{N_2}$ & \\
\end{tabular}
\end{ruledtabular}
\end{table}

Table~\ref{tab:aN2} summarizes the calculations to obtain the nitrogen activity and $\Delta \mu_{Ga}$ under our OMVPE conditions.
The value of $k T \log a_{N_2} = 0.55$~eV at the experimental temperature $T = 1073$ K gives the horizontal coordinate on the phase diagram from Eq.~(\ref{eq:DmuGa}) as $\Delta \mu_{Ga} = -0.40$~eV.
The value of $k T \log p_{NH_3} = -0.30$~eV at the experimental temperature $T = 1073$ K gives the vertical coordinate on the phase diagram from Eq.~(\ref{eq:DmuNH3}) as $\Delta \mu_{NH_3} = -2.40$~eV.
This position is shown on the predicted surface phase diagram, Fig.~\ref{fig:recon_phase}, with a rectangle representing the relatively large uncertainty in $\Delta \mu_{Ga}$.

A recent study of reconstructions on GaN (0001) in the OMVPE environment \cite{2019_Kempisty_PRB100_085304}
included the effects of additional entropy associated with adsorbed species, which leads to a phase diagram that varies somewhat with temperature, even when expressed in chemical potential coordinates.
These effects tend to stabilize reconstructions with H adsorbates at higher $T$, leading to a larger phase field for the 3H(T1) reconstruction than shown in Fig.~\ref{fig:recon_phase}.
This is consistent with our finding that the 3H(T1) reconstruction agrees best with the experimental CTRs for all conditions studied.

\section{Atomic coordinates}

\begin{table} [b]
\caption{ \label{tab:coord_sub} Fractional coordinates of bulk GaN used to calculate the substrate contribution to the CTRs.}
\begin{ruledtabular}
\begin{tabular}{ c | c || r | r | r }
Atom & Site & $x$ & $y$ & $z$\\
$k$ & $n$ & & & \\
\hline
Ga & 1 & 0.5000  &  0.1667 &  -0.5000 \\
Ga & 2 & 0.0000  &  0.6667 &  -0.5000 \\
Ga & 3 & 1.5000  &  0.1667 &  -0.5000 \\
Ga & 4 & 1.0000  &  0.6667 &  -0.5000 \\
Ga & 5 & 0.0000  &  0.0000 &   0.0000 \\
Ga & 6 & 0.5000  &  0.5000 &   0.0000 \\
Ga & 7 & 1.0000  &  0.0000 &   0.0000 \\
Ga & 8 & 1.5000  &  0.5000 &   0.0000 \\
\hline
N & 1 & 0.0000   &  0.0000  &  -0.6232 \\
N & 2 & 0.5000   &  0.5000  &  -0.6232 \\
N & 3 & 1.0000   &  0.0000  &  -0.6232 \\
N & 4 & 1.5000   &  0.5000  &  -0.6232 \\
N & 5 & 0.5000   &  0.1667  &  -0.1232 \\
N & 6 & 0.0000   &  0.6667  &  -0.1232 \\
N & 7 & 1.5000   &  0.1667  &  -0.1232 \\
N & 8 & 1.0000   &  0.6667  &  -0.1232 \\
\end{tabular}
\end{ruledtabular}
\end{table}

\begin{table} 
\caption{\label{tab:coord_recona} Fractional coordinates $x$, $y$, $z$ of atoms in domain $j = 1$ of the 3H(T1) reconstruction used to calculate the $\alpha$ terrace contribution to the CTRs, as well as their differences $\Delta x$, $\Delta y$, $\Delta z$ relative to bulk lattice positions. The differences for H atoms are relative to N sites. The lowest four Ga and N sites are an extra half unit cell of bulk lattice to account for the difference in height of the $\alpha$ and $\beta$ terraces.}
\begin{ruledtabular}
\begin{tabular}{ c | c || r | r | r | r | r | r }
Atom & Site & $x$ & $y$ & $z$ & $\Delta x$ & $\Delta y$ & $\Delta z$\\
$k$ & $n$ & & & & & &\\
\hline
Ga & 1 & 0.5000  &   0.1667  &  -0.5000  &   0.0000  &   0.0000  &   0.0000 \\
Ga & 2 & 0.0000  &   0.6667  &  -0.5000  &   0.0000  &   0.0000  &   0.0000 \\
Ga & 3 & 1.5000  &   0.1667  &  -0.5000  &   0.0000  &   0.0000  &   0.0000 \\
Ga & 4 & 1.0000  &   0.6667  &  -0.5000  &   0.0000  &   0.0000  &   0.0000 \\
Ga & 5 & 0.0000  &   0.0000  &   0.0076  &   0.0000  &   0.0000  &   0.0076 \\
Ga & 6 & 0.5075  &   0.4975  &  -0.0015  &   0.0075  &  -0.0025  &  -0.0015 \\
Ga & 7 & 1.0000  &   0.0050  &  -0.0015  &   0.0000  &   0.0050  &  -0.0015 \\
Ga & 8 & 1.4925  &   0.4975  &  -0.0015  &  -0.0075  &  -0.0025  &  -0.0015 \\
Ga & 9 & 0.4929  &   0.1643  &   0.5223  &  -0.0071  &  -0.0024  &   0.0223 \\
Ga &10 & 0.0000  &   0.6667  &   0.4294  &   0.0000  &   0.0000  &  -0.0706 \\
Ga &11 & 1.5071  &   0.1643  &   0.5223  &   0.0071  &  -0.0024  &   0.0223 \\
Ga &12 & 1.0000  &   0.6714  &   0.5223  &   0.0000  &   0.0047  &   0.0223 \\
\hline
N & 1 & 0.0000  &   0.0000  &  -0.6232  &   0.0000  &   0.0000  &   0.0000 \\
N & 2 & 0.5000  &   0.5000  &  -0.6232  &   0.0000  &   0.0000  &   0.0000 \\
N & 3 & 1.0000  &   0.0000  &  -0.6232  &   0.0000  &   0.0000  &   0.0000 \\
N & 4 & 1.5000  &   0.5000  &  -0.6232  &   0.0000  &   0.0000  &   0.0000 \\
N & 5 & 0.4988  &   0.1663  &  -0.1254  &  -0.0012  &  -0.0004  &  -0.0022 \\
N & 6 & 0.0000  &   0.6667  &  -0.1201  &   0.0000  &   0.0000  &   0.0031 \\
N & 7 & 1.5012  &   0.1663  &  -0.1254  &   0.0012  &  -0.0004  &  -0.0022 \\
N & 8 & 1.0000  &   0.6675  &  -0.1254  &   0.0000  &   0.0008  &  -0.0022 \\
N & 9 & 0.0000  &   0.0000  &   0.3766  &   0.0000  &   0.0000  &  -0.0002 \\
N &10 & 0.5064  &   0.4979  &   0.3775  &   0.0064  &  -0.0021  &   0.0007 \\
N &11 & 1.0000  &   0.0043  &   0.3775  &   0.0000  &   0.0043  &   0.0007 \\
N &12 & 1.4936  &   0.4979  &   0.3775  &  -0.0064  &  -0.0021  &   0.0007 \\
\hline
H &13 & 0.5002  &   0.1667  &   0.8196  &   0.0002  &   0.0000  &  -0.0572 \\
H &15 & 1.4998  &   0.1667  &   0.8196  &  -0.0002  &   0.0001  &  -0.0572 \\
H &16 & 1.0000  &   0.6666  &   0.8196  &   0.0000  &  -0.0001  &  -0.0572 \\

\end{tabular}
\end{ruledtabular}
\end{table}

\begin{table} 
\caption{\label{tab:coord_reconb} Fractional coordinates $x$, $y$, $z$ of atoms in domain $j = 1$ of the 3H(T1) reconstruction used to calculate the $\beta$ terrace contribution to the CTRs, as well as their differences $\Delta x$, $\Delta y$, $\Delta z$ relative to bulk lattice positions. The differences for H atoms are relative to N sites.}
\begin{ruledtabular}
\begin{tabular}{ c | c || r | r | r | r | r | r }
Atom & Site & $x$ & $y$ & $z$ & $\Delta x$ & $\Delta y$ & $\Delta z$\\
$k$ & $n$ & & & & & &\\
\hline
Ga & 1 & 0.5075  &   0.1692  &  -0.5015  &   0.0075  &   0.0025  &  -0.0015 \\
Ga & 2 & 0.0000  &   0.6667  &  -0.4924  &   0.0000  &   0.0000  &   0.0076 \\
Ga & 3 & 1.4925  &   0.1692  &  -0.5015  &  -0.0075  &   0.0025  &  -0.0015 \\
Ga & 4 & 1.0000  &   0.6617  &  -0.5015  &   0.0000  &  -0.0050  &  -0.0015 \\
Ga & 5 & 0.0000  &   0.0000  &  -0.0706  &   0.0000  &   0.0000  &  -0.0706 \\
Ga & 6 & 0.4929  &   0.5024  &   0.0223  &  -0.0071  &   0.0024  &   0.0223 \\
Ga & 7 & 1.0000  &  -0.0047  &   0.0223  &   0.0000  &  -0.0047  &   0.0223 \\
Ga & 8 & 1.5071  &   0.5024  &   0.0223  &   0.0071  &   0.0024  &   0.0223 \\
\hline
N & 1 & 0.0000  &   0.0000  &  -0.6201  &   0.0000  &   0.0000  &   0.0031 \\
N & 2 & 0.4988  &   0.5004  &  -0.6254  &  -0.0012  &   0.0004  &  -0.0022 \\
N & 3 & 1.0000  &  -0.0008  &  -0.6254  &   0.0000  &  -0.0008  &  -0.0022 \\
N & 4 & 1.5012  &   0.5004  &  -0.6254  &   0.0012  &   0.0004  &  -0.0022 \\
N & 5 & 0.5064  &   0.1688  &  -0.1225  &   0.0064  &   0.0021  &   0.0007 \\
N & 6 & 0.0000  &   0.6667  &  -0.1234  &   0.0000  &   0.0000  &  -0.0002 \\
N & 7 & 1.4936  &   0.1688  &  -0.1225  &  -0.0064  &   0.0021  &   0.0007 \\
N & 8 & 1.0000  &   0.6624  &  -0.1225  &   0.0000  &  -0.0043  &   0.0007 \\
\hline
H &10 & 0.5002  &   0.5000  &   0.3196  &   0.0002  &  -0.0000  &  -0.0572 \\
H &11 & 1.0000  &   0.0001  &   0.3196  &   0.0000  &   0.0001  &  -0.0572 \\
H &12 & 1.4998  &   0.4999  &   0.3196  &  -0.0002  &  -0.0001  &  -0.0572 \\
\end{tabular}
\end{ruledtabular}
\end{table}

To provide a detailed example of how we calculate the CTR intensities including the effects of reconstruction, we here provide an example of the atomic coordinates for a particular reconstruction.
The qualitative behavior we observe, that $f_\alpha^{ss}$ increases with growth rate, does not depend upon the reconstruction chosen or the exact values of the atomic coordinates used.
These affect only the precise values of $f_\alpha^{ss}$ obtained, as shown in Table~\ref{tab:table1}.

Tables~\ref{tab:coord_sub}, \ref{tab:coord_recona}, and \ref{tab:coord_reconb} give the atomic coordinates for the 3H(T1) reconstruction obtained in \cite{2012_Walkosz_PRB_85_033308}.
The fractional coordinates $x$, $y$, and $z$ given in the tables are the components of the positions $\mathbf{r}_{kn}$, $\mathbf{r}_{jkn}^\alpha$, and $\mathbf{r}_{jkn}^\beta$ used to calculate the structure factors, normalized to the respective orthohexagonal lattice parameters $a$, $b$, and $c$, i.e. $\mathbf{r} = (ax,by,cz)$.
A $2 \times 2$ surface unit cell is used, equivalent to two orthohexagonal unit cells, so there are 8 Ga and 8 N sites in each.
These coordinates place a bulk Ga site on a $\beta$ layer at the origin.
We use $u = 0.3768$ for the internal lattice parameter of bulk GaN, i.e. the fractional distance between Ga and N sites, which deviates slightly from the ideal $3/8$ value as found in \textit{ab initio} calculations \cite{2012_Walkosz_PRB_85_033308,1999_Stampfl_PRB59_5521} and experiments \cite{2015_Minikayev_XraySpect44_382}.  
Relaxed positions were calculated for a one-unit-cell thick layer at the surface.
For the $\alpha$ terrace, and extra half unit cell of bulk (unrelaxed) atoms is attached to the bottom to account for the difference in height of the $\alpha$ and $\beta$ terraces, as shown in Fig.~\ref{fig:recon}.
Coordinates for only one domain are given.
Those for other 5 domains are obtained by 3-fold rotation about the $6_3$ axis and/or reflection of the $y$ coordinate.
One can see that the Ga atoms bonded to the three adsorbed hydrogens of the 3H(T1) reconstruction relax to higher $z$ positions.

\section{Deposition and evaporation rates}

Under the OMVPE conditions used, we observe that deposition of GaN is Ga transport limited (i.e. the deposition rate is proportional to the TEGa supply rate, nearly independent of $T$ and NH$_3$ supply), and the net growth rate has a negative offset at zero TEGa supply corresponding to an evaporation rate that depends on $T$ and the carrier gas composition (e.g. presence or absence of H$_2$).
To determine the deposition rate for the conditions used in the X-ray study, we used the deposition efficiency (deposition rate per TEGa supply rate) determined from previous studies of CTR oscillations during layer-by-layer growth \cite{2014_Perret_APL105_051602,2019_Ju_NatPhys15_589}.
We also measured the evaporation rates at two higher temperatures and both carrier gas compositions (0\% and 50\% H$_2$), and extrapolated them to the lower temperatures studied here.

\begin{figure}
\includegraphics[width=0.85\linewidth]{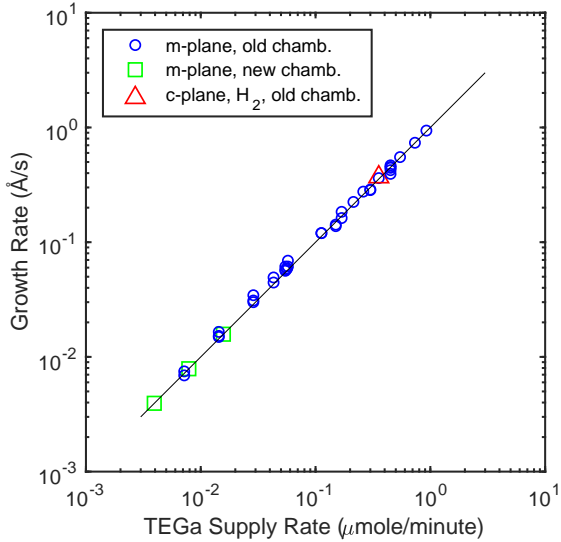}
\caption{Growth rate as a function of TEGa supply determined from CTR oscillations during layer-by-layer growth. Line is fit to new chamber data giving a deposition efficiency of 1.0 (\AA/s)/($\mu$mole/min). \label{fig:g0}}
\end{figure}

\begin{figure}
\includegraphics[width=0.85\linewidth]{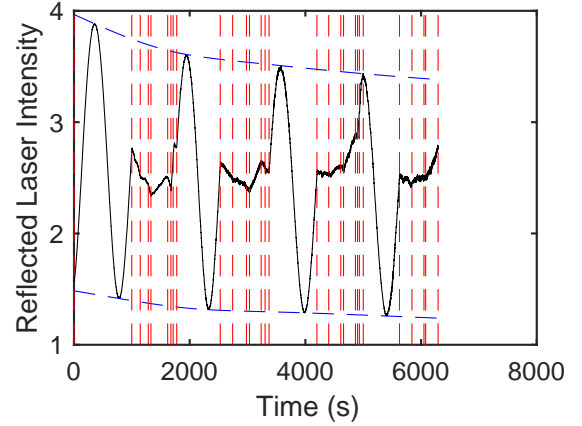}
\caption{Reflected laser signal during growth under various conditions. Vertical dashed lines show times at which conditions changed. \label{fig:g1}}
\end{figure}

Figure~\ref{fig:g0} shows the growth rates measured from CTR oscillations during layer-by-layer growth as a function of TEGa supply \cite{2014_Perret_APL105_051602,2019_Ju_NatPhys15_589}.
In all cases the chamber flows were the same as in the X-ray study reported here (e.g. 2.7 slpm NH$_3$, 267 mbar total pressure).
Almost all data points are for growth on m-plane $(1 0 \overline{1} 0)$ GaN in N$_2$ carrier gas (0\% H$_2$), which exhibits layer-by-layer mode over a wide range of conditions.
Data are shown from both a previous growth chamber (``old'' chamber) \cite{1999_Stephenson_MRSBull24_21} and the current growth chamber (``new'' chamber) \cite{2017_Ju_RSI88_035113,2019_Ju_NatPhys15_589}.
The chambers were designed to have the same flow geometry, and the growth behavior of both appear to be identical.
The data points from the previous chamber range in temperature from $848$~K to $1064$~K; the data points for the current chamber are for $867$~K.
The line shown is a fit to the data from the current chamber, which gives a deposition efficiency of 1.0 (\AA/s)/($\mu$mole/min).
One data point is shown for growth on c-plane (0001) GaN in 50\% N$_2$ + 50\% H$_2$ carrier gas at $900$~K; layer-by-layer growth was only observed on (0001) GaN under this condition.
It agrees with the m-plane data obtained in 0\% H$_2$ carrier, suggesting that the same deposition efficiency can be used for (0001) GaN in either 0\% or 50\% H$_2$ carrier gas.
We expect that there is negligible evaporation at $900$~K in either carrier gas.

\begin{figure}
\includegraphics[width=0.85\linewidth]{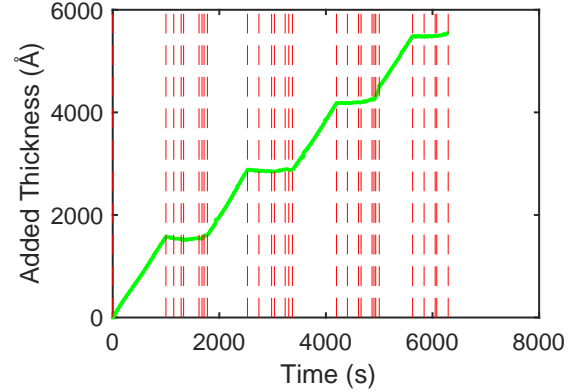}
\caption{Added thickness during growth under various conditions. Vertical dashed lines show times at which conditions changed. \label{fig:g2}}
\end{figure}

\begin{figure}
\includegraphics[width=0.85\linewidth]{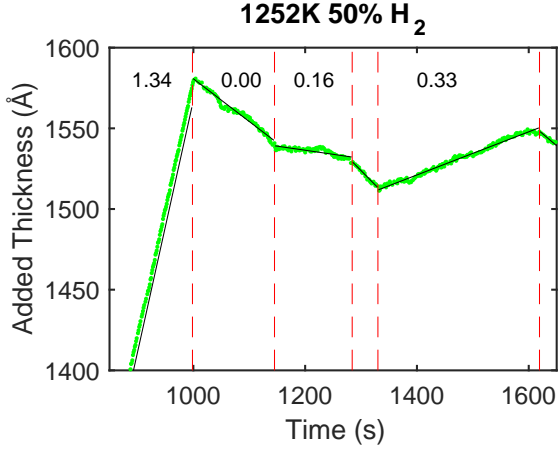}
\caption{Added thickness during growth under TEGa flows shown ($\mu$mol/min) at T = 1252K, 50\% H$_2$. Vertical dashed lines show times at which conditions changed. Black lines show fits to extract net growth rates. \label{fig:g3}}
\end{figure}

\begin{figure}
\includegraphics[width=0.85\linewidth]{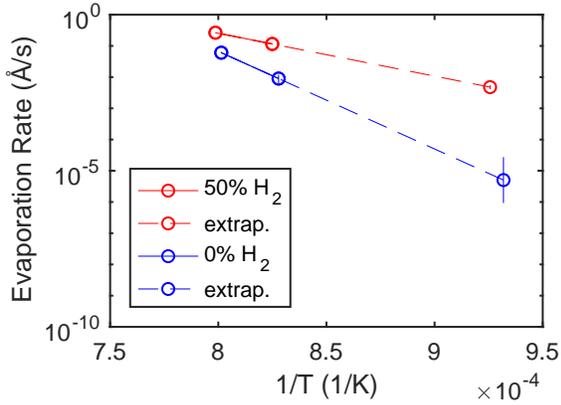}
\caption{Evaporation rate at zero TEGa flow as a function of $T$, with and without H$_2$, with extrapolation to lower $T$. \label{fig:g8}}
\end{figure}

\begin{table}
\caption{ \label{tab:table3} Values of net growth rate $Gc/2$ extracted from laser interferometry measurements for two temperatures and for carrier gas with and without H$_2$, as a function of TEGa flow $f_{TEGa}$. Also shown is fitted $d(Gc/2)/df_{TEGa}$ for each $T$ and carrier gas.} 
\begin{ruledtabular}
\begin{tabular}{ c | c | c || c | c }   
$T$ & H$_2$ & $f_{TEGa}$ & $Gc/2$  & $d(Gc/2)/df_{TEGa}$	\\
(K) & in& ($\mu$mol & (\AA/s) & (\AA/s)/ \\
  & carr. & /min) & & ($\mu$mol/min) \\
\hline
 1208 & 0\%& 0.00 & $-0.009 \pm 0.003$ & $1.19 \pm 0.03$ \\
& & 0.09 & $0.092 \pm 0.003$ & \\
& & 0.33 & $0.322 \pm 0.004$ & \\
& & 1.34 & $1.582 \pm 0.002$ & \\
\hline
 1212 & 50\%& 0.00 & $-0.115 \pm 0.002$ & $1.38 \pm 0.05$ \\
& & 0.09 & $-0.038 \pm 0.002$ & \\
& & 0.33 & $0.248 \pm 0.003$ & \\
& & 1.34 & $1.705 \pm 0.002$ & \\
\hline
 1248 & 0\%& 0.00 & $-0.061 \pm 0.004$ & $1.27 \pm 0.05$ \\
& & 0.16 & $0.042 \pm 0.004$ & \\
& & 0.33 & $0.268 \pm 0.005$ & \\
& & 1.34 & $1.584 \pm 0.002$ & \\
\hline
 1252 & 50\%& 0.00 & $-0.265 \pm 0.004$ & $1.40 \pm 0.03$ \\
& & 0.16 & $-0.050 \pm 0.003$ & \\
& & 0.33 & $0.134 \pm 0.001$ & \\
& & 1.34 & $1.562 \pm 0.001$ & \\
\end{tabular}
\end{ruledtabular}
\end{table}

To determine the evaporation rate at the temperature used in the X-ray study presented here (e.g. $1080$~K), we used laser interferometry to observe the change in thickness of an (0001) GaN film on a sapphire substrate \cite{2017_Ju_RSI88_035113,2005_Koleske_JCrystGrowth279_37}, under conditions of both growth and evaporation at higher $T$.
As the film thickness $d(t)$ changes during growth or evaporation, the back-scattered laser intensity $I(t)$ oscillates with time $t$ due to interference between light reflected from the film surface and the substrate/film interface, according to
\begin{equation}
    I(t) - I_{min} = [I_{max} - I_{min}]
    \left ( \frac{1 + \cos[2 \pi d(t) / d_0]}{2} \right ),
\end{equation}
where $I_{min}(t)$ and $I_{max}(t)$ are the envelope of the minima and maxima, which can vary with time as film roughness changes, and the thickness oscillation period is $d_0 = \lambda/2n$, where $\lambda = 6330$~\AA ~is the wavelength of the light and $n$ is the refractive index of GaN.
This can be inverted to obtain the thickness evolution as
\begin{equation}
    d(t) = \left ( \frac{d_0}{2 \pi} \right ) \cos^{-1} \left ( \frac{2[I(t) - I_{min}(t)]}{I_{max}(t) - I_{min}(t)} - 1 \right ). \label{eq:dt}
\end{equation}
Figure~\ref{fig:g1} shows the evolution of the laser signal with time during the experiment.
We began by growing a full oscillation at a high growth rate to obtain initial values for $I_{min}$ and $I_{max}$.
Once the signal had reach a value intermediate between these limits, where the phase of the oscillation is most accurately determined, we changed the TEGa flow $f_{TEGa}$ to observe the net growth or evaporation rate at some fixed values of $f_{TEGa}$.
Then we changed $T$ and/or the carrier gas concentration, and repeated the process starting with growing a full oscillation at a high rate.
The blue dashed curves in Fig.~\ref{fig:g1} show the interpolated $I_{min}(t)$ and $I_{max}(t)$ envelopes.
Figure~\ref{fig:g2} shows the thickness change with time extracted with Eq.~(\ref{eq:dt}), using a value of $d_0 = 1302$~\AA ~corresponding to $n = 2.431$ \cite{1986_Tapping_JOSAA3_610,1977_Touloulian_TPRCseries13}.
Figure~\ref{fig:g3} shows an expanded region of the thickness evolution, where we varied $f_{TEGa}$ at $T = 1252$~K and 50\% H$_2$ fraction.
The solid lines show linear fits to extract the net growth rate in \AA/s, $Gc/2$, at each value of $f_{TEGa}$.
Similar fits were done for the regions at different $T$ and H$_2$ fraction, and the extracted growth rates are given in Table~\ref{tab:table3}.

We observe that $Gc/2$ becomes negative at $f_{TEGa} = 0$ due to evaporation, and that evaporation is more rapid at higher $T$ and when H$_2$ is present in the carrier gas.
These evaporation rates in 50\% H$_2$ are similar to the rate of $4.2 \times 10^{18}$~m$^{-2}$s$^{-1}$ $= 0.37$~ML/s obtained by \cite{2001_Koleske_JCrystGrowth223_466} at 1300~K with H$_2$ and NH$_3$ at a total pressure of $267$~mbar.
Also shown in Table~\ref{tab:table3} is the deposition efficiency $d(Gc/2)/df_{TEGa}$ obtained from a fit to $Gc/2$ at the four values of $f_{TEGa}$ for each $T$ and H$_2$ fraction.
The values are all similar to but slightly higher than the value of $d(Gc/2)/df_{TEGa} = 1.0$~(\AA/s)/($\mu$mole/min) that we have observed from growth oscillations during layer-by-layer growth at lower $T$, described above \cite{2019_Ju_NatPhys15_589,2014_Perret_APL105_051602}.
The efficiency seems to be slightly larger for 50\% H$_2$ compared with 0\% H$_2$.
This may indicate that the deposition efficiency can vary somewhat as the flow and diffusion fields vary in the chamber with $T$ or carrier gas composition. 

To obtain the evaporation rate at $f_{TEGa} = 0$ at the lower $T$ used in the x-ray experiments reported above, we extrapolated the values for 50\% H$_2$ or 0\% H$_2$ assuming Arrhenius behavior of the evaporation rate, as shown in Fig.~\ref{fig:g8}.
The fitted activation energies are $2.7 \pm 0.1$ and $6.2 \pm 1.2$~eV in 50\% and 0\% H$_2$, respectively.
We obtain evaporation rates of $4.8 \pm 0.8 \times 10^{-3} $~\AA/s at $T = 1080$K with 50\% H$_2$, and $5 \times 10^{-6}$~\AA/s (with error limits of a factor of 5) at $T = 1073$K with 0\% H$_2$.
We have used these evaporation rates, as well as the low-temperature deposition efficiency of $1.0$~(\AA/s)/($\mu$mole/min) and the TEGa flow rates of $0$ or $0.033$ $\mu$mole/min, to calculate the net growth rates in Table~\ref{tab:table1}.


\bibliography{bibliography/2020_GJu_ABsteps_PRLPRB_shortJ}


\end{document}